\documentstyle{mn}
\def\today{\ifcase\month\or
 January\or February\or March\or April\or May\or June\or
 July\or August\or September\or October\or November\or
 December\fi\space\number\day, \number\year}

\def\todmy{\number\day\space\ifcase\month\or
 January\or February\or March\or April\or May\or June\or
 July\or August\or September\or October\or November\or
 December\fi\space\number\year}

\newcommand{\bdisp} {\begin{displaymath}}
\newcommand{\edisp} {\end{displaymath}}
\newcommand{\beqn} {\begin{equation}}
\newcommand{\eeqn} {\end{equation}}
\newcommand{\beqr} {\begin{array}}
\newcommand{\eeqr} {\end{array}}

\newcommand{\tal}{\it et al. \rm}
\newcommand{\AAA}{{A\&A}}

\newcommand{\ApJ}{{ApJ}}

\newcommand{\ApJS}{{ApJS}}
\newcommand{\AJ}{{AJ}}

\newcommand{\MN}{{MNRAS}}

\newcommand{\PASJ}{{PASJ}}

\title{Morphology, photometry and kinematics of $N$-body bars. I Three
models with different halo central concentrations} 
\author[E. Athanassoula, A. Misiriotis]
       {
       E. Athanassoula$^{1}$, A. Misiriotis$^{1,2}$ \\
$^1$ Observatoire de Marseille, 
2 Place Le Verrier, 
F-13248 Marseille Cedex 4, France \\
$^2$ University of Crete, Physics Department, P.O. Box 2208, 710 03
   Heraklion, Crete, Greece \\}

\date{Accepted .
      Received ;
      }

\pagerange{\pageref{firstpage}--\pageref{lastpage}}
\pubyear{2000}

\begin{document}

\maketitle

\label{firstpage} 
\begin{abstract}
We discuss the morphology, photometry and kinematics of the bars which
have formed in three
$N$-body simulations. These have initially the same disc and the same
halo-to-disc mass 
ratio, but their haloes have very different central concentrations. The 
third model includes a bulge. The bar in the model with the centrally 
concentrated halo (model MH) is much stronger, longer and thinner than 
the bar in the model with the less centrally concentrated halo (model MD).
Its shape, when viewed side-on, evolves from boxy to peanut and then to
X-shaped, as opposed to that of model MD, which stays boxy.
The projected density profiles obtained from cuts along the bar major axis,
both for the face-on and the edge-on views, show a flat part, as 
opposed to those of model MD which are falling rapidly. A Fourier
analysis of the face-on density distribution of model MH shows very
large $m$ = 2, 4, 6 and 8 components. Contrary to this for model MD
the components $m$ = 6 and 8 are negligible.
The velocity field of model MH shows strong deviations from axial symmetry, 
and in particular has wavy isovelocities near the end of the bar when 
viewed along the bar minor axis. When viewed edge-on, it shows cylindrical 
rotation, which the MD model does not. The properties of the bar of the model 
with a bulge and a non-centrally concentrated halo (MDB) are
intermediate between 
those of the bar of the other two models. All three models exhibit a lot of 
inflow of the disc material during their evolution, so that by the end
of the simulations the 
disc dominates over the halo in the inner parts, even for model MH,
for which the halo and disc contributions were initially comparable in
that region.

\end{abstract}


\section{Introduction}
\indent

A bar is an elongated concentration of matter in the central parts of
a disc galaxy. Within this loose and somewhat vague definition fit a
number of very different objects. Thus different bars have very
different masses, axial ratios, shapes, mass and colour
distributions. They can also have widely different kinematics. Several
observational studies have been devoted to the 
structural properties of bars and/or to their morphology,
photometry and kinematics, thus providing valuable information on these
objects and on their properties. 

Many $N$-body simulations of the evolution of disc galaxies have
witnessed the formation of bars. Most studies have focused on
understanding what 
favours or hinders bar formation. Not much work, however, has been
done on the ``observable'' properties of $N$-body bars. This is quite
unfortunate since such studies are necessary for the comparison
of real and numerical bars. In fact several observational studies have
taken an $N$-body simulation available in the literature and have
analysed it in a way similar to that used for the observations in
order to make comparisons (e.g. Kormendy 1983, Ohta, Hamabe \&
Wakamatsu 1990, L\"utticke, Dettmar \& Pohlen 2000). Although this is
very useful, it suffers from lack of generality, since the specific
simulation may not be appropriate for the observational question at
hand, and since it does not give a sufficient overview of the
alternative properties $N$-body bars can have. Here we will approach
the comparisons between real and $N$-body bars from the simulation side,
giving as wide a range of 
alternatives as possible, while making an analysis as near as possible
to that used by observers. We hope that in this way our work will be of use
to future observational studies and will provide results for detailed
comparisons. 

An obvious problem when comparing $N$-body bars to real bars is that
simulations trace mass, while observations give information on the
distribution of light. The usual way to 
overcome this hurdle is to assume a constant $M/L$ ratio. This
assumption should be adequate for the inner parts of galaxies
(e.g. Kent 1986, Peletier \& Balcells 1996),
particularly early types that have relatively little star
formation, and in the NIR wavelengths, where the absorption
from dust is least pronounced. 

By their nature, real bars can be observed in only a much more limited
way than $N$-body bars. Galaxies are projected on the plane of the sky
and their 
deprojection is not unambiguous, particularly for barred galaxies,
which are the object of the present study. This problem of course does
not exist for simulations, which we furthermore can ``observe'' from
any angle we wish. It is thus possible to ``observe'' the same
snapshot both face-on and edge-on. This is of course impossible to do
for real galaxies and has led to a number of complications
e.g. regarding the nature of peanuts and the 3D structure of bars. A
second limitation is that in observations light is integrated along
the line of sight and one can not observe the various components
separately, as in $N$-body bars. Finally the biggest limitation comes
from the fact that there is no direct way to observe dark matter,
while in $N$-body simulations the halo can be analysed as any other
component of the galaxy. All these limitations lead to complications in
the comparisons, but are also one of the
reasons for which the observations of $N$-body bars  are most
useful. We can observe our bars both in the restricted manner
that real galaxies allow and in the more detailed manner accessible to
simulations and, by comparing the two, derive the signatures of the latter
in the former. This can help us obtain information on
properties of real bars which are not directly observable.

In this paper we will discuss at length the observable properties of
three simulations. In section~\ref{sec:models} we present the simulations
and their initial conditions. In section~\ref{sec:basic} we present the
basic properties of our three fiducial models and in section~\ref{sec:ellips}
the shape and the axial ratio of the isodensities in the bar
region. Projected density profiles with the disc seen face-on and edge-on are
presented in sections~\ref{sec:faceon} and \ref{sec:edgeon},  
respectively. In section~\ref{sec:mcomp} we present the Fourier components 
of the mass distribution seen face-on, in section~\ref{sec:barlength}
we compare various ways of measuring the bar length, and in
section~\ref{sec:peanut} we quantify the peanut shape. Kinematics are
presented in sections~\ref{sec:rot} and \ref{sec:disp} and the shape of the
bulge is discussed in section~\ref{sec:bulge}. We summarise in
section~\ref{sec:summary}. In a companion paper (hereafter Paper II)
one of us (E.A.) will present more 
simulations, compare to observations and discuss implications about the
distribution of the dark matter in barred galaxies.
 
In this paper and its companion we will willingly refrain
from discussing 
resonances, their location and their effect on the evolution of the
bar. This discussion necessarily implies the knowledge of the
pattern speed of the bar, which is not directly available from
observations. Together with the discussions relying on 
dynamics and/or some knowledge of the orbital structure it will be
left for a future paper.

\section{Simulations }
\label{sec:models}
\indent

\begin{figure*}
\includegraphics{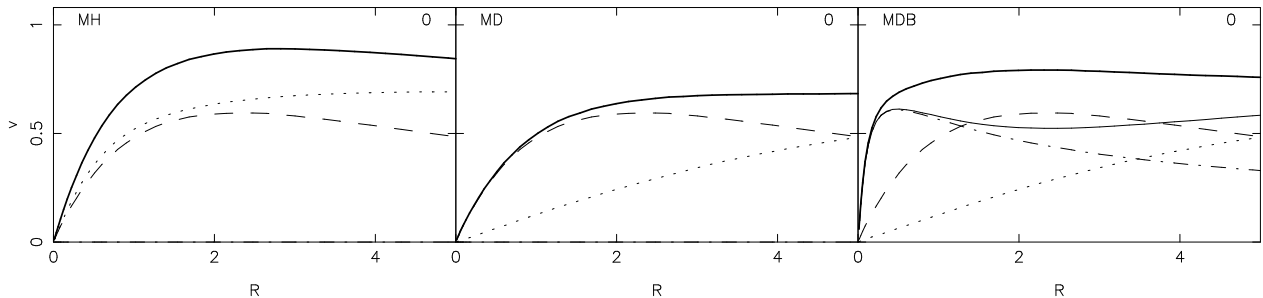}
\vspace{6.0cm}
\caption{Circular
velocity curves of our three fiducial models at the beginning of the
simulation. The dashed, dotted 
and dash-dotted lines give the contributions of the disc, halo
and bulge respectively, while the thick full lines give the total circular
velocity curves. For simulation MDB we also give the the total
contribution from the two spherical components with a thin solid line.
The left panel corresponds to simulation MH, the middle one to simulation MD and
the right one to simulation MDB. The simulation name is given in the
upper left corner and the time 
in the upper right corner of each panel.}
\label{fig:rotcur0}
\end{figure*}

We have made a large number of simulations of bar-unstable discs, three
of which we will discuss in this paper. Each is characteristic of
a class of models, other members of which will be discussed in paper II.

In order to prepare the initial conditions we basically followed the
method of Hernquist (1993), to which we brought 
a few improvements, described in Appendix~\ref{sec:initdetail}.

The density distribution of the disc is given by

\begin{equation}
\rho_d (R, z) = \frac {M_d}{4 \pi h^2 z_0}~~exp (- R/h)~~sech^2 (\frac{z}{z_0}),
\end{equation}

\noindent
that of the bulge by

\begin{equation}
\rho_b (r) = \frac {M_b}{2\pi a^2}~~\frac {1}{r(1+r/a)^3},
\end{equation}

\noindent
and that of the halo by

\begin{equation}
\rho_h (r) = \frac {M_h}{2\pi^{3/2}}~~ \frac{\alpha}{r_c} ~~\frac {exp(-r^2/r_c^2)}{r^2+\gamma^2}.
\end{equation}

\noindent
In the above $r$ is the radius, $R$ is the cylindrical radius, $M_d$,
$M_b$ and $M_h$ are the masses of the disc, bulge 
and halo respectively, $h$ is the disc radial scale length, $z_0$ is
the disc vertical scale thickness, $a$ is the bulge scale length,
and $\gamma$ and $r_c$ are halo scale lengths.
The parameter $\alpha$ in the halo density equation is a normalisation
constant defined by 

\begin{equation}
\alpha = [1 - \sqrt \pi~~exp (q^2)~~(1 -erf (q))]^{-1}
\end{equation}

\noindent
where $q=\gamma / r_c$ (cf. Hernquist 1993). In all simulations we
have taken $M_d$ = 1, $h$ = 1 and have represented the disc with
200\,000 particles. The halo mass, calculated to infinity, is
taken equal to 5, and $r_c$ is always 
taken equal to 10. The halo mass distribution is truncated at 15. The
disc distribution is cut vertically at $z_{cut} = 3 z_0$ 
and radially at half the halo truncation radius, i.e. $R_{cut} =
7.5$. The velocity distributions are as described by 
Hernquist (1993) and in Appendix~\ref{sec:initdetail}. 

The first two fiducial models that we will discuss at length here have very
different central concentrations. For the first one we have taken
$\gamma$ = 0.5, so that the halo is
centrally concentrated and in the inner parts has a contribution
somewhat larger than that of the disc. Since the
mass of the disc particles is the same as that of the halo particles,
the number of particles in the halo is set by the mass of the halo
within the truncation radius (in this case roughly 4.8) and in this simulation
is roughly equal to 963030. We will 
hereafter call this model ``massive halo'' model, or, for short,
MH. For the second model we have taken $\gamma$ = 5, so that the disc
dominates in the inner parts. The halo is represented by 931206
particles. We will 
hereafter call this model ``massive disc'' model, or, for short, MD.
Both MH and MD models have no bulge. In order to examine the effect of
the bulge we will consider a third fiducial model, which is similar to MD
but has a bulge of mass $M_b = 0.6$ and of scale length $a =
0.4$. We will hereafter refer to this model as ``massive disc 
with bulge'', or, for short, MDB.
In the three fiducial simulations we adopted a disc thickness of
$z_0$~=~0.2 and $Q$ = 0.9. 
Their circular velocity curves are shown in Figure~\ref{fig:rotcur0},
together with the contribution of each
component separately. For model MDB we also show the total
contribution from the two spherical components. The halo and disc
contributions in the inner parts are comparable in the case of model
MH, while the disc dominates in 
model MD. For model MDB as one moves from the center outwards one
has first a bulge dominated part, then a disc dominated part and
finally a halo dominated part. Thus its evolution could in principle
be different from both that of the MH and that of the MD models. All
three cases have a flat rotation curve, at least within a radius of
five disc scale lengths. If we consider larger radii, say up to 15
disc scale lengths, then the rotation curves for models MD and MDB are
still flat, while that of model MH decreases, because its halo is
centrally concentrated. By adding an extra extended halo component we
can keep the rotation curve flat up to such distances. This, however,
more than doubles the total mass of the system, and raises the number
of particles accordingly, to a number which was beyond our CPU capacities,
particularly seen the large number of MH-type simulations described in
this and the companion paper. We thus ran a simulation with particles
of double the mass, and therefore half the number. This allowed us to
check that the introduction of this extended halo does not change
qualitatively the results of the morphology, photometry and kinematics 
of the bar. Having established this, and since this simulation has
only 100\,000 particles in the disc, so that the noise is higher and
the quantities describing the bar which are discussed here less 
well defined, we will present in this paper
the results of the simulation without the more extended halo component.

These three simulations are part of a bigger ensemble, covering 
a large fraction of the
available parameter space. In this paper we will discuss only these
three, which are each characteristic of a certain type of
simulations. By limiting ourselves to three simulations we will be able
to make a very thorough analysis of each case. In Paper II we
will discuss more simulations in order to assess how certain parameters,
like $Q$ or the thickness of the disc, influence the main results
presented here.

The simulations were carried out on a Marseille
Observatory GRAPE-5 system consisting of two
GRAPE-5 cards (Kawai \tal 2000) coupled via a PCI interface (Kawai
\tal 1997) to a Compaq DS 20 or an XP-1000.
We used a GRAPE treecode similar to that initially
built for the GRAPE-3 system (see e.g. Athanassoula \tal 1998). For
simulation MH we used an
opening angle of 0.7, while for simulations MD and MDB 
we used an opening angle of 0.6. With the XP-1000 front end, one time-step
for $10^6$ particles takes roughly 15 sec for $\theta$ = 0.6
and less than 12 sec for $\theta$ = 0.7. We used a
softening of 0.0625 and a time-step  of 0.015625. This gave us an
energy conservation better than or of the order of one part 
in a thousand over the entire simulations, which were terminated after
$t$ = 900, i.e. after 57\,600 time-steps. Full information on all the particles
in the simulation is kept every 20 time units, while information on the 
total potential and kinetic energy and on the center of mass of the system 
is saved every 8 time steps, i.e. every 0.125 time units. We calculate
on line (Athanassoula \tal 1998) the 
amplitude and the phase of the $m$ = 2 and 4 Fourier components of the 
mass every 0.5 time units. We also produce gif files of the face-on
and edge-on  
distributions of the disc particles every 0.5 time units which, when viewed 
consecutively as a movie with the help of the xanim software, give a
good global view of the evolution.

In this paper, unless otherwise noted, all quantities are given in
computer units scaled so that 
the scale length of the disc is unity, the total mass of the disc
equal to 1 and $G$~=~1. It is easy to convert them to standard astronomical
units by assigning a mass and a scale length to the disc. Thus, if the
mass of the disc is taken to be equal to  5 $\times$ $10^{10}$
$M_{\odot}$ and its scale length equal to 3.5 kpc, we find that the unit of
mass is 5 $\times$ $10^{10}$ $M_{\odot}$, the unit of length is 3.5
kpc, the unit of velocity is 248 km/sec 
and the unit of time is 1.4 $\times$ $10^7$ yrs. 
Thus time 500 corresponds to 7 $\times$ $10^9$ yrs and time 
800 to 1.1 $\times$ $10^{10}$ yrs.
This calibration, however, is not unique. Adopting different values
for the disc scale length and mass would have led to alternative
calibrations.

\section{Two types of $N$-body bars }
\label{sec:basic}
\indent

\begin{figure*}
\includegraphics{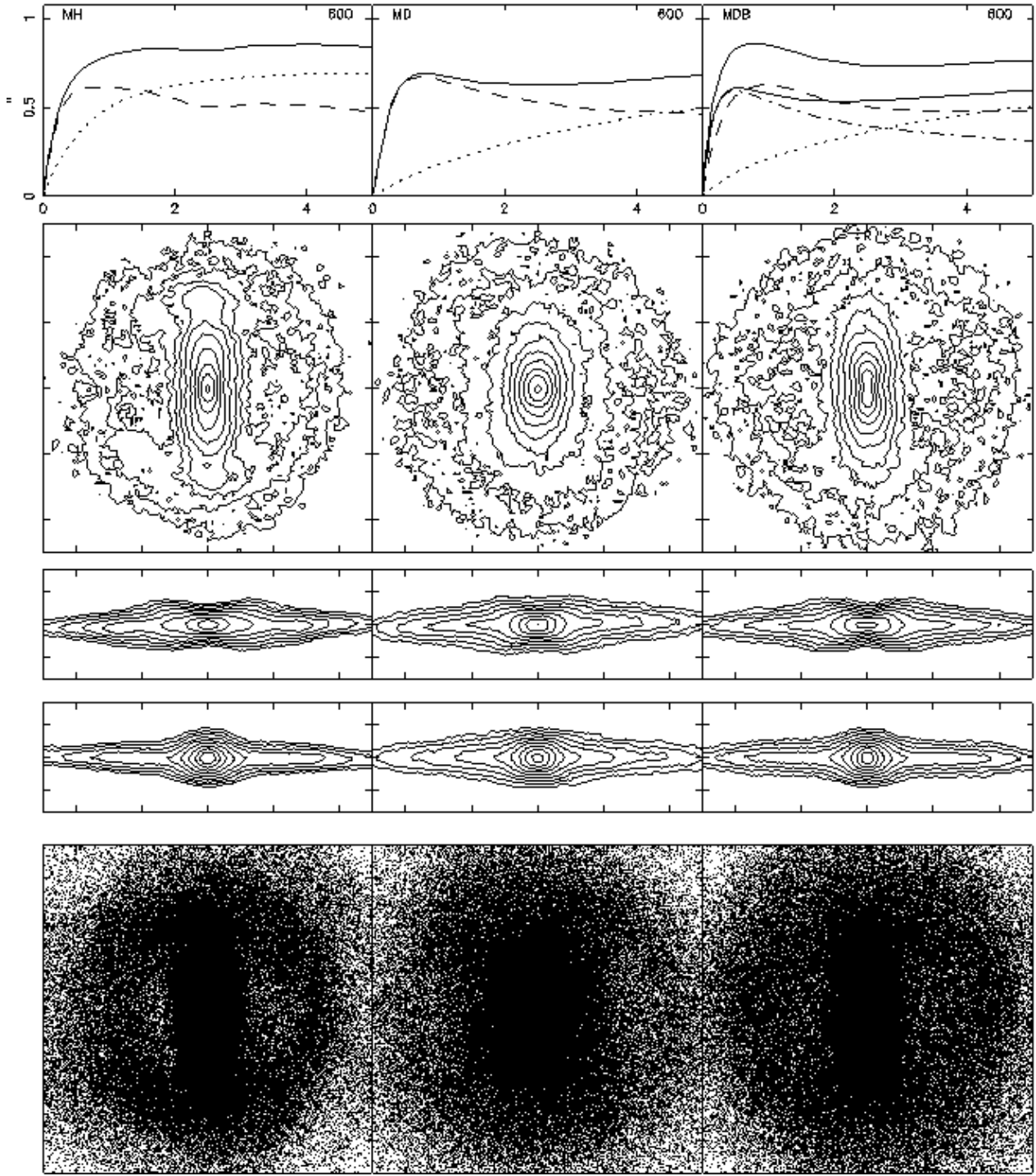}
\vspace{20.cm}
\caption{Basic information on the three fiducial simulations at time
$t$ = 600. Left
panels correspond to simulation MH, middle ones to simulation MD and
right ones to simulation MDB. The upper panels give the circular
velocity curve. The dashed, dotted
and dash-dotted lines give the contributions of the disc, halo
and bulge respectively, while the thick full lines give the total circular
velocity curves. For simulation MDB we also give the the total
contribution from the two spherical components (thin solid
line). The second row of panels gives the isocontours of the
density of the disc particles projected face-on,
and the third and fourth row the side-on and
end-on edge-on views, respectively. The fifth row of panels gives the
dot-plots of the particles in the $(x,y)$ plane. The side of the box
for the face-on views is 10 units, and the height of the box for the
edge-on views is 3.33 units.
}
\label{fig:basic600}
\end{figure*}

\begin{figure*}
\includegraphics{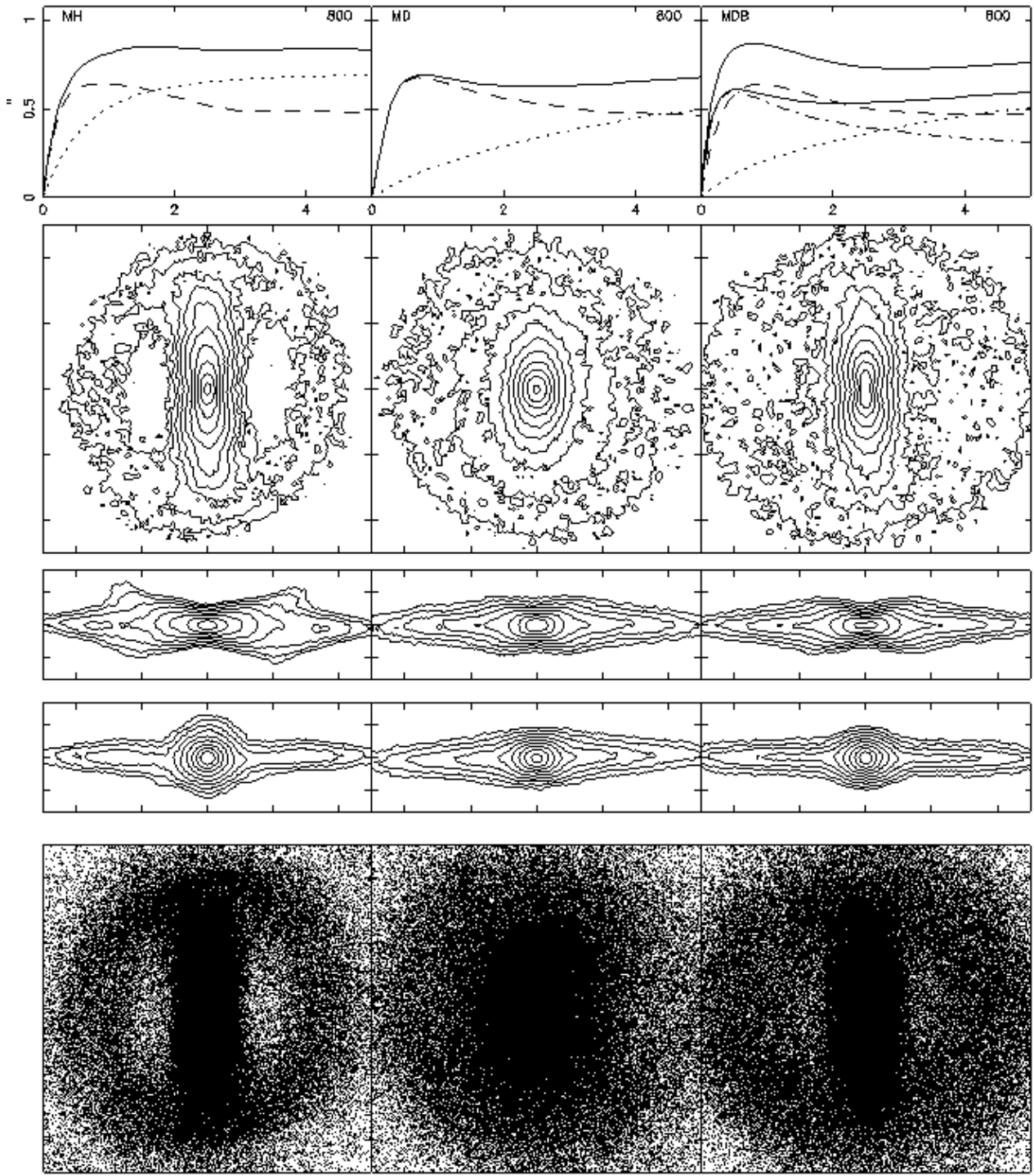}
\vspace{20.cm}
\caption{Same as the previous figure, but for time $t$ = 800.}
\label{fig:basic800}
\end{figure*}

Figures~\ref{fig:basic600} and \ref{fig:basic800} give some basic
information on the three fiducial simulations at times 600 and 800 
respectively. The upper panels give the total circular
velocity curves together with the contribution of each component
separately. In all three cases the disc material has moved inwards 
as a result of the evolution and the
configurations become much more centrally concentrated. As a result,
the circular velocity curve of model MH rises much faster than
initially and after the rising part
stays roughly constant. Also the center-most part is disc dominated,
contrary to what was the case at the start of the simulation. The
circular velocity curves of models MD and MDB develop a peak near the
center, also due to the increased central concentration of the disc
material. In model MD all the central part is disc dominated, as it 
was at the start of the simulation. On the other hand for model MDB there is
no region which is disc dominated because of the joint effect of the
halo and bulge. There is not much difference between the circular 
velocity curves at times 600 and 800.

Observed face-on{\footnote{Unless otherwise noted, in this paper we
adopt a coordinate system such that the $x$ and $y$ axes lie on the
equatorial plane, the $y$ axis being along the bar major axis.}}
(second and fifth rows in Figures~\ref{fig:basic600} and 
\ref{fig:basic800}) the disc
particle distributions are quite different in the three fiducial
cases. The bar in the MH model is longer and thinner than the bar in 
model MD. At time $t$ = 600 it has ``ansae'' at the
extremities of the bar, 
similar to those observed in early type barred galaxies. These
structures do not exist at time 800. The bar in model MD is
rather short and fat, while the bar in
model MDB is intermediate both in length and shape of the bars of
models MD and MH, but nearer to that of MH.

Model MH has a ring, which observers 
would call an inner ring since it surrounds the bar and since its radius is
roughly equal to the bar semi-major axis. It shows up clearer in
the dot-plots of the fifth row than in the isodensity plots of the
second row. It can be discerned in the isodensity plots at time 800, 
but not at time 600, 
due to the fact that the region between the ring and the bar has a much
lower projected surface density at the later time. Such a ring
does not exist in the MD model. For model MDB it shows 
as a broad diffuse structure. In model MH the ring has a density
enhancement near the ends of the bar both at times 600 and 800, which 
is slightly stronger towards the leading side. In fact
rings have not been witnessed before 
this in purely  stellar $N$-body simulations, with the notable exception
of the fiducial simulation of Debattista \& Sellwood (2000). It is worth
noting that both in their simulation and in
our MH model the maximum of the halo rotation curve
is near the center of the galaxy.

We use the methods described in Appendix~\ref{sec:techniques} to
determine the parameters of the ring. Applying the local 
(global) method to model MH at time $t$ = 600 we find that
the ring has a radius of 3.0 (3.2) and an axial ratio $b/a$ of
0.7 (0.8). It is very thick -- the width of the fitting gaussian 
(cf. Appendix~\ref{sec:techniques}) being of the order of 1.8 (1.4) 
--, it is aligned 
parallel to the bar major axis and its mass is 31 (28)
percent of the total disc mass. For time $t$ = 800 we find that
the radius is 3.6 (3.7) and the axial ratio 
0.8 (0.9). The width of the fitting gaussian is 1.4 (1.3), 
it is aligned 
parallel to the bar major axis and its mass is 20 (20)
percent of the total disc mass. This shows that the ring  has become
less eccentric and that its diameter has increased with time. 
The results from the two methods agree well, but they both
overestimate the mass of the ring, since they 
consider the total mass under the fitted gaussian, so that the wings 
of the gaussian contribute substantially. Had we truncated the gaussian 
we would have obtained a considerably lower value. Since, however, we do 
not know where to truncate we leave it as is, and let the reader obtain
the mass after truncation from the values of the parameters we give
above. 

Model MDB also shows a ring. At times 600
its radius is 3.1 (3.2) and its axial ratio 0.9 (0.9). 
It is even broader than the ring in model MH -- the width of the
gaussian being of the order of 2.1 (1.9) -- and thus we can not give a
reliable estimate of its mass, as argued in Appendix~\ref{sec:techniques}.
At times 800 its radius is 3.6 (3.6) and its shape is near-circular.
Again we note that the size of the ring has increased with time and
that its shape has become more circular. Again there is good agreement
between the two methods.

The edge-on projections of the disc density (given by
isodensities in 
the third and fourth rows in Figures~\ref{fig:basic600} and
\ref{fig:basic800}) are also widely
different in the three fiducial cases. In the side-on{\footnote {We 
call side-on view the edge-on view for which the line of sight is 
along the bar minor axis. Similarly we call end-on view the edge-on 
view where the line of sight is along the bar major axis.}} view, model MD 
has a boxy outline. On the other hand model MH is peanut-like
at time 600 and is X shaped at time 800. Model MDB has the form of a
peanut. Seen end-on, models MH and MDB show a central
structure of considerable size, which could well be 
mistaken for a bulge. This is particularly strong for model MH at
time 800. At that time we also see the signature of the ring on the
side-on view, as closing isodensities on either side of the central 
area. 

\section{The axial ratio and shape of the isodensities in the bar 
region. Face-on view }
\label{sec:ellips}

\begin{figure*}
\includegraphics{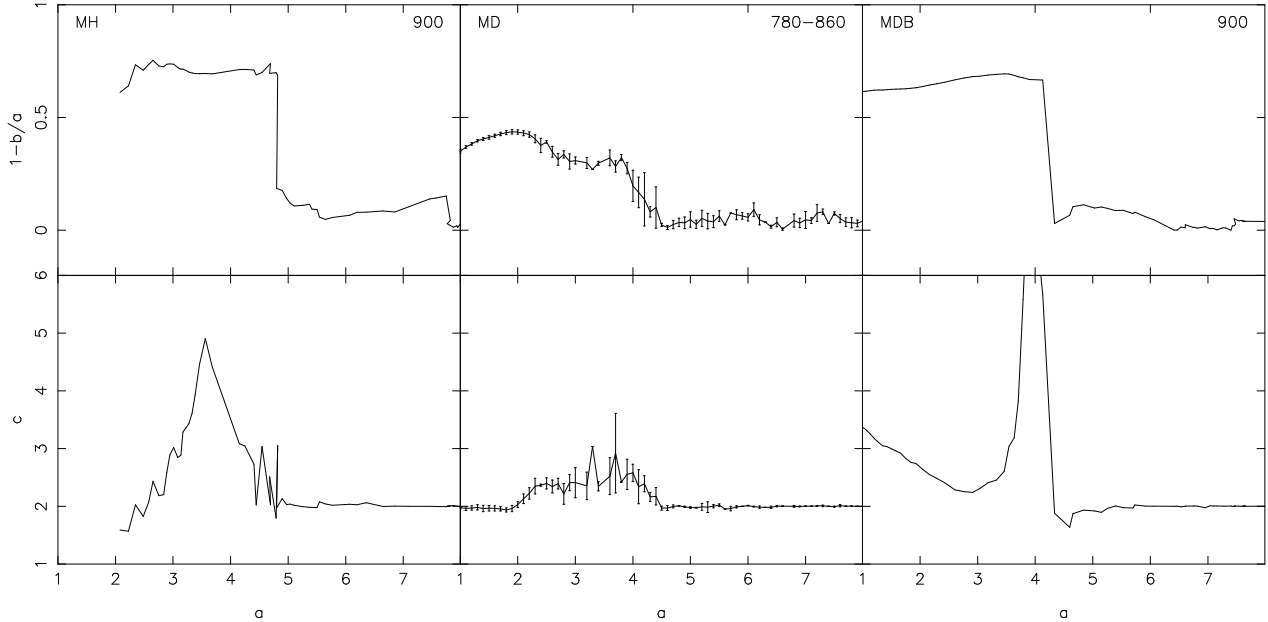}
\vspace{8.6cm}
\caption{The upper panels show the run of the ellipticity 1 - $b/a$ as
a function of the semi-major axis $a$. The lower panels show the run of
the shape parameter $c$, also as a function of $a$. The left panels
corresponds to model MH, the middle ones to model MD and the right ones
to model MDB. For reasons explained in the text we show for models MH
and MDB the results at a given time, namely $t$~=~900, while for model 
MD we give the
average of a time interval, namely [780,~860]. The dispersion during 
that time is indicated by the error bars. The simulation 
name is given in the upper left corner and the times
in the upper right corner of the upper panels.
}
\label{fig:axrat}
\end{figure*}

In order to measure the axial ratio and shape of the bar isodensities
seen face-on 
we project all disc particles on the $(x, y)$ plane on which we
superpose a  200 X 200 cartesian grid covering a square of length
$(-16,16)$. The density at the
center of each cell is calculated by counting the number of particles
in the cell and dividing by its area. The density at intermediate
points is calculated using bilinear interpolation. We then 
fit generalised ellipses to the isodensities. The equation of the
generalised ellipse, initially introduced by Athanassoula \tal (1990),
is

\begin{equation}
(|x|/a)^c + (|y|/b)^c = 1,
\end{equation}

\noindent
where $a$ and $b$ are the semi-major and semi-minor axes respectively,
and $c$ is a parameter describing the shape of the generalised
ellipse. For $c$ = 2 we obtain a standard ellipse, for $c < 2$ a
lozenge, while for $c > 2$ the shape approaches a rectangle, and will, for
simplicity, be called hereafter rectangular-like. Athanassoula \tal (1990) have
used generalised ellipses to quantify the shape of the
isophotes of strongly barred early-type galaxies and we
will use them here to describe the shape of our $N$-body
bars. Unfortunately the comparison between the two is not as
straightforward as it sounds. All galaxies in the Athanassoula \tal
sample have a sizeable bulge, so that it was not possible to get
information on the isophotes of the inner parts of the bar. Since
isophotes are determined both by the light of the bulge and that of
the bar, it was necessary to blank out the inner parts of the galaxy,
where the bulge light dominates. In a few cases there were also small
spiral features starting off from the end of the bar, and these also
had to be blanked out. Since the sample was relatively small it was
possible to determine interactively the areas to be blanked out, separately
in each frame.

Models MH and MD have no bulge, so it is possible to continue the fits
much further in than in the case of early-type barred galaxies. (It is
nevertheless recommended to exclude the innermost few pixels, where
the shape of the pixel may influence the shape of the isophote). 
Comparisons with observations, though, will not extend all the way to
the center. Model MDB has a sizeable bulge, but we fitted the
generalised ellipses to the disc component only. We did this in order
to be able to discuss the effect of the bulge on the real shape of the
bar, rather than the shape of the disc and bulge components combined. 
This is trivial to do in the case of models, but of course
impossible for real galaxies. We will discuss at the end of this
section how the existence of the bulge obscures the issue when the
generalised ellipses are fitted to both components, as is the case in
real galaxies.

For every simulation and for every time step for which full
information on the particle position is available -- i.e. 45 time
steps per simulation -- we fitted
generalised ellipses to 70 isodensities covering the density
range in the disc. We repeated the exercise twice, once generously excluding a
central region, and the other excluding only a very small central
region. For reasons of homogeneity and seen the very large number of
frames to be treated, we  
determined the area to be blanked out automatically and not
interactively. For this we first took the difference of 
the projected density along the minor and major axes of the bar and
calculated the radius at which this is maximum. In the first passage
we excluded all the region within this radius, and in the second the
region within one tenth of this radius. We made several tests with
other blanking radii and thus asserted that our results are not dependent on
this radius. Since the blanking out was done automatically we did not
blank out the spirals coming from the ends of the bar. This presents
problems in the early steps of the simulations, where the spirals are
important and therefore influence wrongly the fits. In the later
stages though, no such spirals are present, and the fits pertain to the
bar only.
 
For models MD the generalised ellipses
fit the isodensities very well all the way to the center, and thus
the results obtained with the generous and with the limited blanking
agree very well in the region where both give information. This is 
true also for most of the MDB and a large fraction of the MH cases.
For some MH cases, however, the $b/a$ values 
obtained with the two fits are
in good agreement, but the $c$ values are not. The reason is that the
generalised ellipse is too simple a shape to fit properly the isophotes which
have ansae or rectangular-like tips of the bar. In fact, for
those cases, the shape 
enclosed by the isodensities is fatter around the major axis than what
the generalised ellipse will allow. Thus the $c$ value obtained by 
such a fit is not meaningful if the central parts have not been blanked 
out. This shortcoming
was not clear for real galaxies, since there the inner parts are bulge
dominated and thus were blanked out. 

The upper panels of Figure~\ref{fig:axrat} show the ellipticity,
$1~-~b/a$, as a function of semi-major axis $a$ for our three fiducial
models. For model MD we present the average of the five times in the
time interval [780,~860] for which we have full information on all the
particle positions. This was done in order to get a better
signal-to-nose ratio, and was possible because the axial ratio does
not show any clear evolution with time. This is not the case for
models MH and MDB, where evolution is present and we can not make
averages without loosing information. These curves establish
quantitatively the general impression we had already from
Figures~\ref{fig:basic600} and \ref{fig:basic800}, namely that the bar
in model MH is much 
thinner than that in model MD, the one in MDB being
intermediate. Indeed the $(1 - b/a)_{max}$ is 0.75, 0.44 and
0.69 respectively for models MH, MD and MDB. The run of the ellipticity
with radius is also different in the three cases. For model MD the 
$1~-~b/a$ rises to a maximum, occurring around $a$ = 
1.9, and then drops again. In many cases the drop after the maximum is
more gradual than that shown in the Figure. This means that the
isodensities are nearer 
to circular in the innermost and outermost parts of the bar and are
more elongated in the intermediate region. For models MH and 
MDB the ellipticity curve is quite
flat, meaning that the bar is thin even in the inner
and outermost parts. After this flat region there is a very steep drop to a
near-circular value. The value of $a$ at which this drop occurs increases
noticeably with time, so that making time averages would have smoothed out
this feature considerably. Right 
after the steep drop there is a region with very  
low values of the ellipticity, after which the ellipticity rises again, 
albeit to a lower value. This intermediate low 
ellipticity region corresponds to the near-circular isophotes in the 
ring region, while the disc outside it has more elongated isophotes, 
their orientation being perpendicular to the bar, as expected. The 
disc of models MD and MDB outside the bar region has less elongated 
isophotes. 

The lower panels of Figure~\ref{fig:axrat} give, for our three fiducial
models, the shape parameter $c$ as a function of the semi-major axis $a$. 
This is less well determined than the axial ratio. One reason
is that the value of $c$ is much more sensitive to noise than the
axial ratio, and our isodensities have significant noise since we have
``only'' 200 000 particles in the disc. A second reason, as argued 
above, is that in
cases where the shape of the isodensity is not well described by a
generalised ellipse the parameter $c$ is more prone to error than the
axial ratio $b/a$. Although individual profiles can show a number of
spurious jumps, due mainly to noise, it is nevertheless possible to see 
that there are considerable differences between the three fiducial
models. The form of this curve for model MD does
not evolve with time, so we give here, as in the upper panel,
the average of the five 
times in the time range [780,860], in order to increase the signal-to-noise
ratio. We note that the value of $c$ increases from a value around 2 in
the center-most parts to a value around 3 for $a$ = 3.3 and then
drops again to a value near 2. This means that the isophotes
are well fitted by ellipses in the center-most and outer parts of the
bar, with a region in between with somewhat rectangular-like
isophotes. The deviations from the elliptical shape, however, are never
very pronounced. 
The shape of the $c$ profile for model MH is very different. 
It has a sharp peak in the outer
parts of the bar, after which the value of $c$ drops abruptly. The
maximum value is very high, around 5, but the region in which high
values are seen is rather narrow. The value of $a$ at which the $c$
value drops sharply increases considerably with time, and for this
reason we have not taken time averages, but present in
Figure~\ref{fig:axrat} only the values for $t$ = 900. For model 
MDB the fits to generalised ellipses can extend to the 
innermost regions. Starting from the innermost parts we find first
an extended region where $c$ drops from over 3 to roughly 2, and then
a very sharp peak of $c$~=~8.1 at a radius of 3.9. The drop after the
maximum is very steep. For model MDB, as for model MH, 
the value of $a$ at which the maximum occurs increases with time.
 
We repeated the exercise for model MDB, this time keeping both the disc and
the bulge particles, in order to see what effect the inclusion of the
bulge would have on the results. We find that the shape parameter $c$
has considerably smaller values, in fact in between 2 and 3. 
Also the ellipticity drops, by something
of the order of 0.1. Of course this value will depend on the density profile of
the bulge, but the general trend should always be that including the
bulge will make the isophotes nearer to ellipses and less eccentric.

The isophotes in the outermost disc region are also of interest, since they
are often used by observers to deproject a galaxy, with the
assumption that they are circular. In order to check whether this is a
reasonable hypothesis we fitted ellipses to the outermost parts of the
disc of our
three fiducial models. We find that the deviations of the ellipticity
from unity - i.e. of the isodensities from circularity - are less than
0.1 in the 
outermost parts. For model MH the slight elongation is perpendicular to the bar
major axis, while for models MD and MDB it is along it. This may mean
that for the latter two we are outside the outer Lindblad resonance,
and for the former not. We will discuss this further in a future paper,
after we have introduced the measurements of the pattern
speed. Independent of the orientation of the outermost ellipses, their
ellipticity is very near unity. Our fiducial models thus argue that it
is reasonable to 
use the outermost disc isophotes to deproject an observed barred galaxy.

\section{Density profiles along the major and minor axes of bars seen face-on
}
\label{sec:faceon}      

\begin{figure*}
\includegraphics{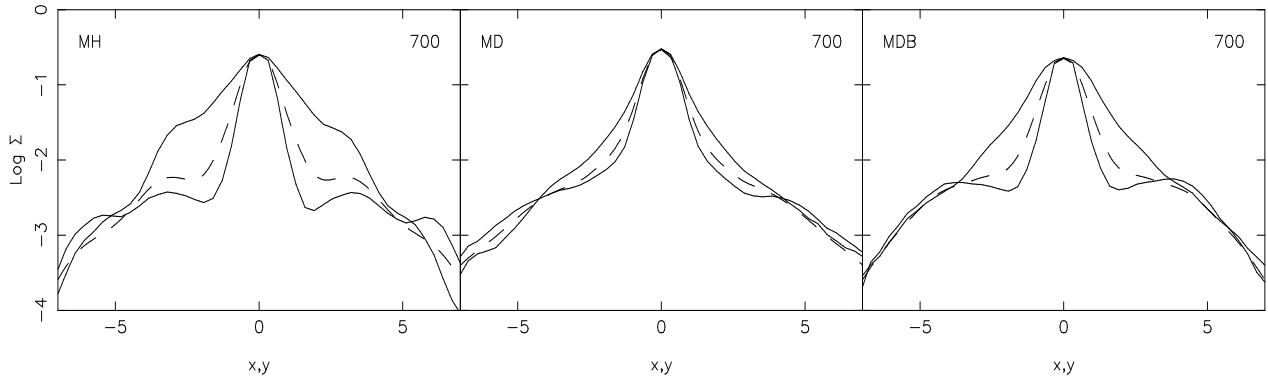}
\vspace{5.0cm}
\caption{
Projected density profiles along the bar major and minor axes (solid lines)
and azimuthally averaged (dashed lines). The disc is seen face-on. The
left panel corresponds to model MH, the middle one to model MD and the
right one to model MDB. The simulation name is given in the upper left 
corner and the time in the upper right corner of each panel. }
\label{fig:faceprof}
\end{figure*}

Figure~\ref{fig:faceprof} shows the projected density profiles along
the minor and 
major axes of the bar of our three fiducial simulations at time $t$ =
700. We note that the three sets of profiles
differ significantly. For MD models the bar has along its major axis a 
fast decreasing profile, with a slope that is steeper than that of the
outer disc. The MDB major axis profile resembles that of MD. On the 
other hand for MH models the profile has a flat part followed by
a relatively abrupt drop at the end of the bar region.
MD bar profiles stay of the same shape all through the simulation, while 
the shape of MH profiles shows some evolution; their
flatness shows clearest between times 540 and 740.

Model MH also shows a considerable density concentration in the 
innermost parts, well above the flat part. If such an enhancement was
present in observations it would be attributed to a bulge component
and this would have been wrong since we know that this simulation
does not have a bulge. It is just an important central concentration 
of the disc component. It can also be present in models MD and MDB.
Since, however, the bar has an exponential-like profile, it is
not easy to disentangle the contribution of this component from 
that of the remaining bar. In this respect it is worth noting that the central 
projected surface density has roughly the same value for the
three models.

The minor axis profiles of MD and MH models fall at
considerably different rates. This, however, is not a new finding. It
is just due to the fact that the MH bars are considerably
thinner than the bars of type MD. In fact from the major and minor axis
cuts alone one can get an estimate of the axial ratio of the bar at
different isophotal levels, simply by drawing horizontal lines at
given isophotal levels on plots
like those of Figure~\ref{fig:faceprof} and measuring the radii at
which the two profiles reach the given isophotal level. 

\section{Density profiles of bars seen edge-on}
\label{sec:edgeon}

\begin{figure*}
\includegraphics{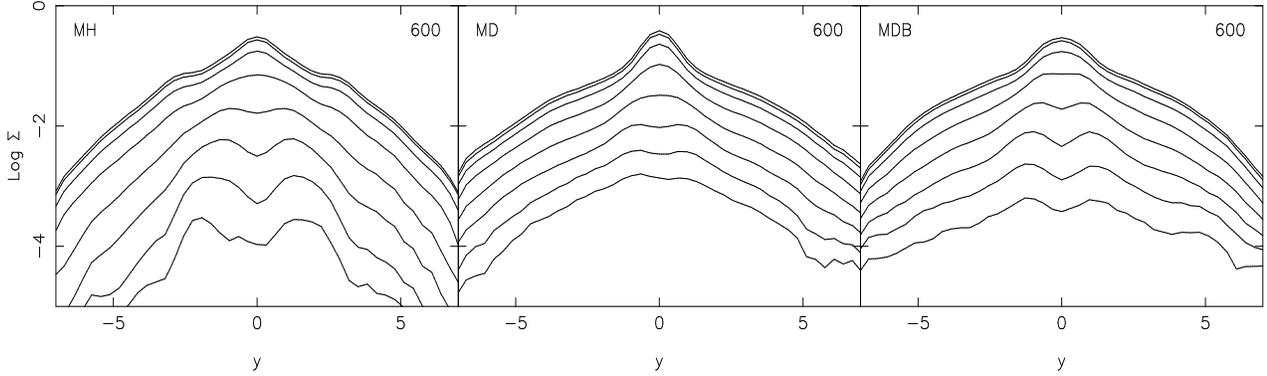}
\vspace{5.0cm}
\caption{
Projected surface density along cuts which are parallel to the major axis (the
bar is seen side-on) at equidistant 
$z$ values differing by $\Delta z$~=~0.2. The uppermost curve corresponds 
to a cut at $z$ = 0 and the lower-most one to a cut at $z$ = 1.4. The
left panel corresponds to model MH, the middle one to model MD and the
right one to model MDB, all taken at time 600. The simulation name is given 
in the upper left corner and the time in the upper right corner of each
panel.
}
\label{fig:zcuts}
\end{figure*}

Let us now observe our three fiducial models edge-on with the bar seen
side-on, i.e. with the line of sight perpendicular to the bar major axis.
Figure~\ref{fig:zcuts} shows the projected density along cuts parallel 
to the major axis made at
equidistant $z$ values differing by $\Delta z$ = 0.2. Again we note
clear differences between the three models. For cuts which are
offset from the equatorial plane of the galaxy, model MH shows a clear
minimum at the center, followed on either side by a maximum, followed by a
steep drop. This is the direct signature of the peanut, which is due to
the minimum thickness in the center, followed by two maxima on either
side. This is very clear in the three cuts which correspond
to the highest $z$ displacements, i.e. for $z$ between 1.0 and
1.4. For models MD and MDB the peanut
signature is much less pronounced. The profiles have, if any, a very shallow
minimum at the center followed by a maximum nearer to the center than
in model MH, visible in a rather  
restricted range of $z$ displacements. 

Another clear distinction between the MH and MD models is the form of the
profile at $z$ = 0. For model MH it shows clear ledges on either side
of the peak, while there are no corresponding structures on the MD and MDB
profiles. Thus for model MH at time 500 the ledge at $z$ = 0 
ends roughly at 2.9. These ledges can be found also
on cuts somewhat 
displaced from $z$ = 0, but disappear after the displacement has
become too large, in this case $|z|$ of the order of 0.4. Similar values 
are found for time 800. The ratio
of the maximum distance to which the ledge on the $z$ = 0 cut extends 
to the radius of
the very steep drop found on cuts offset from $z$ = 0 is roughly 1.5.
 
The peak in the center is due to material that has accumulated in the
central regions of the galaxy and is the edge-on analog of the corresponding 
density enhancement seen in the face-on profiles . This peak can be
seen only on profiles near $z$ = 0, and disappears once the
displacement from the equatorial plane becomes sufficiently large, i.e. 
$|z| > 0.8$. 

\section{Fourier components of the face-on density distribution }
\label{sec:mcomp}

\begin{figure*}
\includegraphics{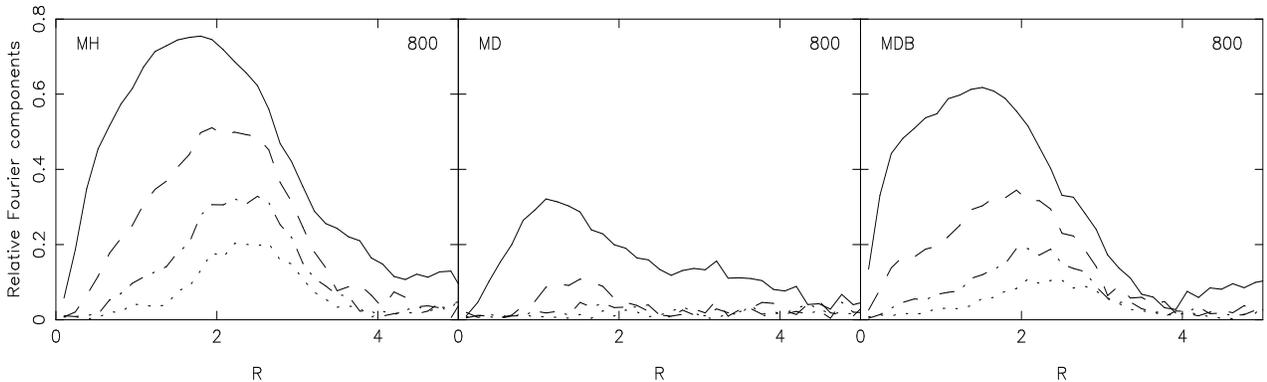}
\vspace{5.0cm}
\caption{Relative
amplitude of the $m$ = 2 (solid line), 4 (dashed line), 6 (dot-dashed 
line) and 8 (dotted line) components of the mass or density. The
left panel corresponds to model MH, the middle one to model MD and the
right one to model MDB. The simulation number is given in the upper
left corner and the time in the upper right corner of each panel.
}
\label{fig:mcomp}
\end{figure*}

Figure~\ref{fig:mcomp} shows the relative amplitude of the $m$ = 2, 4, 6 and 8
components of the mass or density for our three fiducial cases. To
calculate them we first projected the disc particles on the $(x,y)$
plane and then obtained the Fourier components 

\begin{equation}
A_m (r) = \frac {1}{\pi} \int _0 ^{2 \pi} \Sigma (r, \theta) cos (m \theta) d \theta, ~~~~~~~m=0, 1, 2, .....
\end{equation}

\noindent
and

\begin{equation}
B_m (r) = \frac {1}{\pi} \int _0 ^{2 \pi} \Sigma (r, \theta) sin (m \theta) d \theta, ~~~~~~~m=1, 2, .....
\end{equation}

\noindent
by dividing the surface into annuli of equal width $\Delta r$ = 0.14 and
calculating the $A_m$ and $B_m$ for each annulus. In practice instead of the
surface density $\Sigma (r, \theta)$ we use the mass, but this comes
to the same thing since we will be using only the ratios of the
amplitudes $\sqrt{A_m^2+B_m^2}/ A_0$.
 
The maximum amplitude of the $m$ = 2
component is biggest in the MH and smallest in the MD model,
reflecting the fact that the bar in the MH model is the strongest and
that in the MD model is the weakest of the three. Also in model MH 
the maximum occurs at a larger radius than in MD, and in general the
region where this component is large extends to larger radii. This
reflects the fact that, as seen in section~\ref{sec:basic}, the bar in
model MH is the longest and that in MD the shortest of the
three. The most striking difference, however, between the three models
concerns the higher order terms. Thus for the MD and MDB the $m$ = 6
and $m$ = 8 components basically stay within the noise, 
while in the MH model they are 44 and 27
percent of the $m$ = 2 respectively. The relative importance of the
$m$ = 4 is also very different. It is 68, 34 and 55  percent of the
$m$~=~2 respectively in models MH, MD and MDB. 

At time 500 the form of the curves is roughly the same as at
800. There is, however, a growth of the relative amplitude with time, 
which is quite strong for simulations MH, and exists also for MDB. 
Thus the maximum relative amplitude at time 500 is 0.56, 0.32 and 0.52 for 
models MH, MD and MDB respectively. Also the locations of the maxima
of simulations MH and MDB move
further out with time. At time 500 they are at 
$R$ = 1.4, 1.4 and 1.2 respectively for the three models. Finally 
the secondary maximum occurring in simulation MH at  
larger radii -- $R$ somewhat less than 4 -- is very 
pronounced at time 500 and is located at a smaller radius.

\section{Length of the bar }
\label{sec:barlength}

Measuring the length of the bar is not unambiguous and several methods
have been proposed so far. In order to apply them to our simulations 
we have had in certain cases to extend
them or modify. Thus we have defined the length of the bar as follows:

\begin{enumerate}
\item From the value of the semi-major axis at which the ellipticity
is maximum ($L_{b/a}$). One can use this value as such, or take a
multiple of it. Here we adopted the former.

\item From the steep drop in the run of the ellipticity (or axial
ratio) as a function of the semi-major axis ($L_{drop}$).  
As we saw in section \ref{sec:ellips}, in model MH the ellipticity 
presents a steep drop towards the end of the bar
region. Although this is not the case for model MD we will
still use the radius at which the axial ratio shows the largest drop
as a possible measure of the bar length, because it is a
straightforward and direct method and it does not introduce any {\it ad hoc}
constants. It could prove to be a good estimate for cases like model MH,
where the drop is clear. It can unfortunately not be used blindly,
since in some cases, as in model MD, there may not be a steep drop.

\item From the phase of the bar ($L_{phase}$). The phase of a perfect
(theoretical) bar 
should be constant. This is of course only approximately true in
$N$-body or observed bars. Nevertheless the phase varies little 
and the bar can be defined as the region within
which the phase varies less than a given amount. Instead of considering
the differences between the phases in two consecutive annuli, which
may be heavily influenced by noise, we use a somewhat less local
definition. We first calculate the phase of the $m$ = 2 component of
the whole disc, i.e. the phase of the bar. Then we
repeat the exercise after slicing the disc in circular annuli. 
In the innermost parts the amplitude of the $m$~=~2 is very low 
and the noise in the phase can therefore be important. Then there is a
region where the phase is nearly constant and equal to the phase of
the bar, and then it starts varying with radius. We define as
length of the bar the radius of the first annulus for which the phase
differs by more than $asin (0.3)$ from the phase of the bar. The choice of the
constant is of course {\it ad hoc}. It has simply been estimated so as to
give reasonable results in a few test cases.

\item From the $m$ = 2 component, or from the ratio of the $m$~=~2 to
the $m$ = 0 components ($L_{m=2}$). In a model in which
the disc does  
not respond to the bar, one would expect the end of the 
bar to be where all components except $m$ = 0 go to zero. This is not
true in $N$-body
simulations and in real galaxies, where the disc responds to the bar
and thus is not perfectly  
axisymmetric. One can, nevertheless, get an 
estimate of the bar length from the radius at which the relative $m$ = 2 
component is less than a given fraction of
the maximum, provided there is no clear spiral structure. Here we 
will adopt as length of the bar the radius at
which the relative $m$ = 2 component reaches 20 per cent of its
maximum value. Again the choice of the constant is {\it ad hoc}, estimated so as to
give reasonable results in a few test cases.

\item From the face-on profiles ($L_{prof}$). 
We take the difference between the projected density profiles along
the bar major and minor axes. This is of course zero at the center and 
increases with distance to reach a maximum and then drops. In a 
theoretical case of a bar in a rigid disc the end of the bar would 
be where the two projected density profiles became again equal. Since
in $N$-body simulations the disc is responsive, the difference will
not be zero even in the disc. We thus define as bar length the outer
distance from the center at which the difference falls to 5 per cent of the
maximum. Again the choice of the constant is {\it ad hoc}, estimated so as to
give reasonable results in a few test cases.
   
\item From the edge-on profiles ($L_{zprof}$). We can define as 
length of the bar the distance of the end of the ledge on the 
$z$ = 0 cut from the center of the galaxy.

\item From the ratio of the intensities in the bar and the inter-bar
region ($L_{Ohta}$ and $L_{Agr}$). Ohta \tal (1990) defined as bar
region the zone with a contrast 
$I_b/I_{ib}$ exceeding 2, where $I_b$ and $I_{ib}$ are, respectively,
the bar and the inter-bar intensities. These can be simply defined as 
$I_b = I_0 + I_2 + I_4 + I_6$ and $I_{ib} = I_0 - I_2 + I_4 -
I_6$. This criterion was modified by  Aguerri \tal (2000) to delineate the
region where $I_b/I_{ib} > 0.5 [(I_b/I_{ib})_{max} -
(I_b/I_{ib})_{min}] + (I_b/I_{ib})_{min}$. The length of the bar is
then simply the outer radius at which  $I_b/I_{ib} = 0.5 [(I_b/I_{ib})_{max} -
(I_b/I_{ib})_{min}] + (I_b/I_{ib})_{min}$.
This definition can be applied to $N$-body bars by changing the intensity
for the density, so we will adopt it as one of our
definitions. Contrary to other definitions it has
the advantage of being applicable to analytic models, for which the
length of the bar is known exactly. We have thus applied it to all
models in Athanassoula (1992) and found in all cases an agreement of
better than 4 per cent, except for the models with the very thin
homogeneous bars, of axial ratio $a/b$ larger than 4, where the error
can reach 8 per cent.
\end{enumerate}

Several of the above definitions are much easier to
apply in $N$-body models 
than in real galaxies, since in the former we are sure to be in the
equatorial plane of the disc, and thus we do not have the considerable
uncertainties which a deprojection from the plane of the
sky can bring to real galaxies. Most of them have been used in one study or
another. Unfortunately they have never been applied all to the same
case, so as to allow comparisons. For this reason we have applied them
all to all simulations  and to all time steps, and give the results
for two times of our three fiducial simulations in
Table~\ref{tab:barlength}. Since 
these estimates suffer from the existence of noise we have used an
average over a given time range. This time range, however, should be rather
small, since the length of the bar increases with time. Thus the
$\sigma$ given in the table is an overestimate of the real uncertainty,
since it includes the effect of time evolution. The first
column in Table~\ref{tab:barlength} gives the name of the model, the
second one the time range over which the average was taken and columns
3 to 9 give the estimates of the various methods. The second line in
each case is similar, but contains the values of the dispersion. For
the first two estimates ($L_{b/a}$ and $L_{drop}$) we use the ellipse
fitting with the generous blanking (cf. section~\ref{sec:ellips}), since
this is more generally applicable. For the
cases where the ellipticity is almost 
flat (like MH and MDB) $L_{b/a}$ is not meaningful, while for MD the
ellipticity profile does not show a steep drop, so that 
$L_{drop}$ is not well defined. Generally the largest values are
given by $L_{phase}$ and $L_{prof}$, and the
smallest values are given by $L_{b/a}$ and $L_{Agr}$. $L_{drop}$ and $L_{m=2}$
give generally estimates which are intermediate. 

\begin{figure*}
\includegraphics{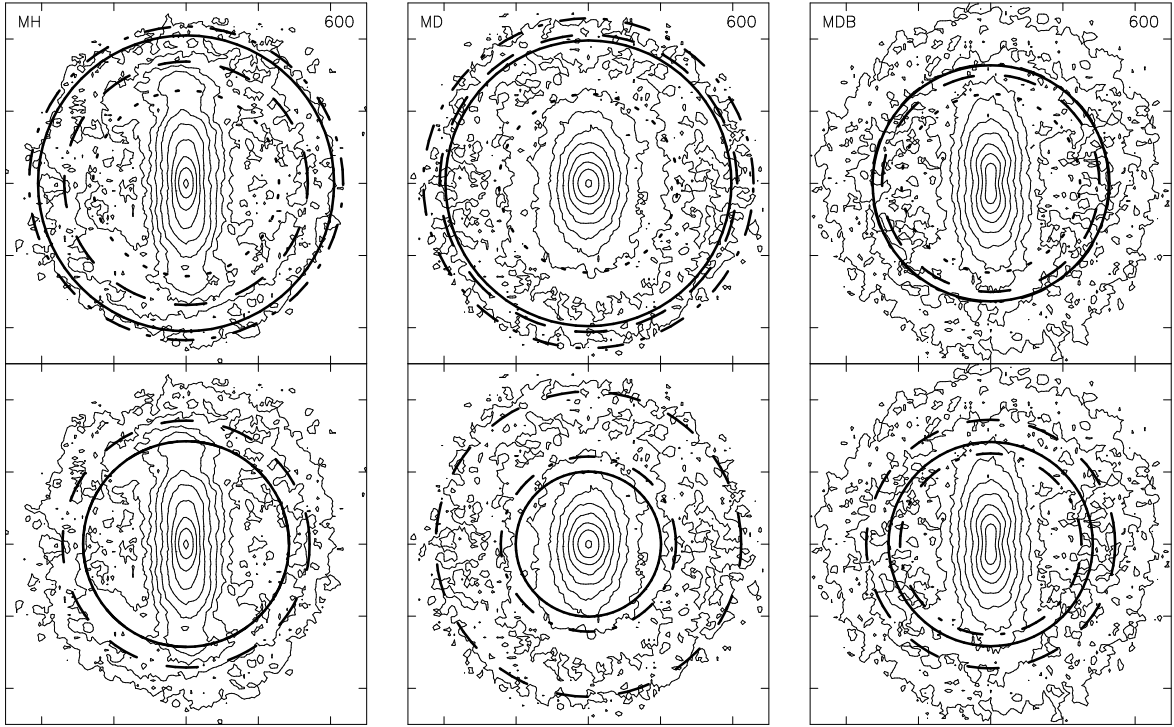}
\vspace{11.cm}
\caption{
The main determinations of the bar length superposed on isocontours
of the projected density for the disc component. In the upper panel we
give $L_{phase}$ (solid line), $L_{m=2}$ (dashed), $L_{prof}$
(dash-dotted) and $L_{Agr}$ (dotted). In the lower panel we give
$L_{b/a}$ (solid line), $L_{drop}$ (dashed) and $L_{zprof}$
(dash-dotted). The name of the simulation is
given in the upper left corner of the upper panels. The isocontours
correspond to the time given in the upper right corner. The bar length
estimates are obtained by averaging the results of seven times,
centered around the time given, and spaced at equal intervals of
$\delta t$ = 20, as in Table 1. The length of the tick marks is 2 
computer units.
}
\label{fig:lb600}
\end{figure*}

\begin{figure*}
\includegraphics{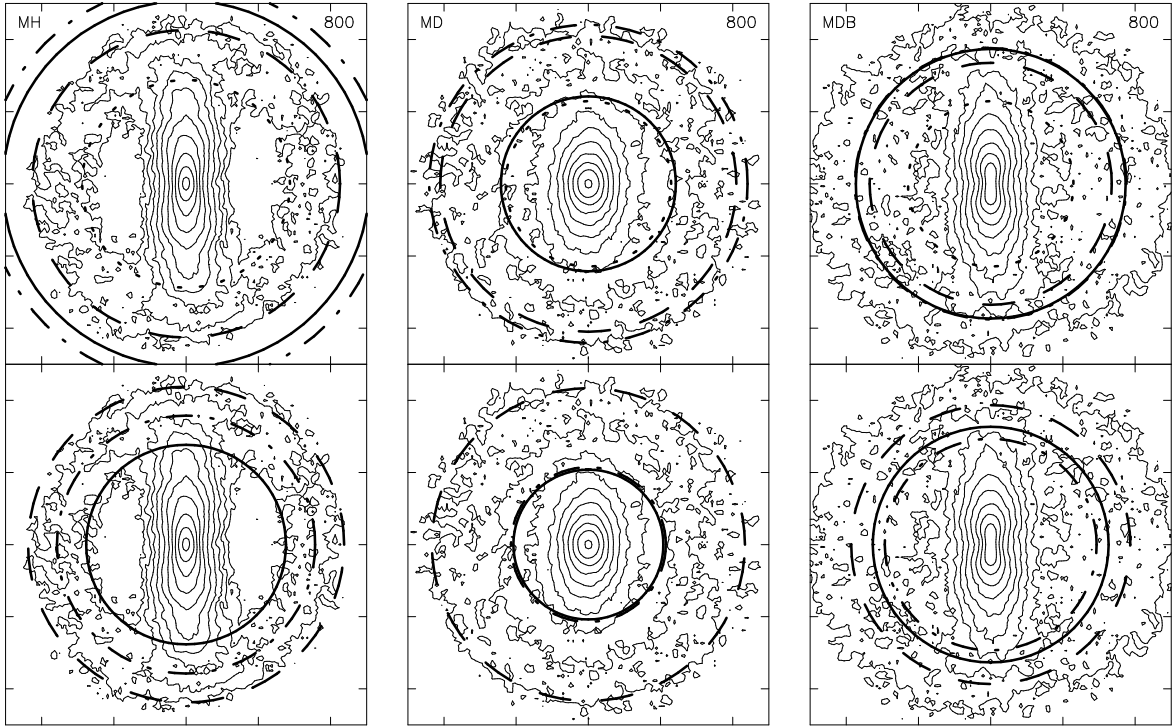}
\vspace{11.cm}
\caption{
Same as the previous figure, but for time 800.
}
\label{fig:lb800}
\end{figure*}

It is clear that the differences between the various methods are larger
than the dispersions within each method. It is thus of interest to
see which method, if any, gives the best results. For this in
figures~\ref{fig:lb600} and \ref{fig:lb800} we overlay on the
isodensity curves taken at times 600 and 800 respectively, circles
with radii the various estimates of the bar length. We note that there
is no single method which fares well in all cases. For time 600 the
best estimates are $L_{drop}$ and $L_{m=2}$ for MH, $L_{zprof}$ and
$L_{Agr}$ for MD, and $L_{b/a}$ and $L_{m=2}$ for MDB. For time 800 the
best estimates are $L_{zprof}$ for MH, $L_{phase}$ for MD, and
$L_{drop}$, $L_{phase}$ and $L_{prof}$ for MDB. 

It is also interesting to see whether the average values represent the
length of the bar well or not.
Column 10 gives the average of all the estimates ($L_1$) in the first line
and the standard deviation in the second. We also tried a second
average ($L_2$) for which we omitted the entries which were considered
unsafe. For this we rely on the applicability of the method to the
models and not on whether they give results compatible with the visual
estimates of the bar. 
For cases MH and MDB $L_{b/a}$ is meaningless, since the ellipticity
profile is very flat. Similarly for model MD $L_{drop}$ is 
unreliable since there is no clear drop in the $b/a$ profile. For
model MD there are two more estimates that could be unreliable. By
examining the individual face-on profiles we see that the difference between
the profile along the bar and perpendicular to it is not very dependent
on radius (cf. Figure~\ref{fig:faceprof}) and thus $L_{prof}$ is badly
defined. Finally the ledge is also not clear on the edge-on profiles,
so that the $L_{zprof}$ estimate is also unreliable. We thus omitted
these two estimates as well for model MD{\footnote{In fact the ledge
is also difficult to see for model MDB up to time 620. So a more
careful treatment would omit this estimate for part of simulation MDB.}}.
$L_2$ and its standard deviation are given in the first
and second line of column 11. Taking into account the error
bars, we see that average values are good estimates of the length
of the bar. This, however, is not a big help, since the error bars are
rather large, particularly so for model MD.

Columns 12 to 15 contain information on the peanut length and will be
discussed in the next section.

\begin{table*}
\caption[]{Length of the bar}
\begin{flushleft}
\label{tab:barlength}
\begin{tabular}{lllllllllllllll}
\hline
model & Time & $L_{b/a}$ & $L_{drop}$ & $L_{phase}$ & $L_{m=2}$ &
$L_{prof}$ &$L_{zprof}$ & $L_{Agr}$ & $L_1$ & $L_2$ &
$LP_1$ & $LP_2$ & $LP/L_2$ & \\
\hline\noalign{\smallskip}
\noalign{\smallskip}
MH & 540-660 & 2.9 & 3.4 & 4.1 & 3.4 & 4.3 & 2.9 & 2.6 & 3.4 & 3.4 & 1.7 & 1.9 & 0.5 \\
MH & 540-660 & 0.2 & 0.6 & 0.8 & 0.2 & 0.5 & 0.2 & 0.3 & 0.7 & 0.7 & 0.3 & 0.2 &  \\
\noalign{\smallskip} 
MD & 540-660 & 2.0 & 3.5 & 4.0 & 4.1 & 4.6 & 2.4 & 2.5 & 3.3 & 3.1 & 1.0 & 1.1 & 0.3 \\
MD & 540-660 & 0.1 & 1.6 & 1.6 & 0.2 & 0.3 & 0.5 & 0.2 & 1.0 & 1.1 & 0.1 & 0.1 & \\
\noalign{\smallskip} 
MDB & 540-660 & 2.8 & 3.4 & 3.3 & 3.0 & 3.3 & 2.5 & 2.3 & 3.0 & 3.0 & 1.3 & 1.4 & 0.5 \\
MDB & 540-660 & 0.2 & 0.1 & 0.1 & 0.1 & 0.1 & 0.2 & 0.1 & 0.4 & 0.5 & 0.3 & 0.1 & \\
\noalign{\smallskip} 
MH & 740-860 & 2.8 & 4.4 & 5.1 & 4.3 & 5.5 & 3.6 & 2.9 & 4.1 & 4.3 & 2.2 & 2.5 & 0.6 \\
MH & 740-860 & 0.2 & 0.1 & 0.9 & 0.4 & 0.5 & 0.1 & 0.1 & 1.0 & 1.0 & 0.0 & 0.1 &  \\
\noalign{\smallskip} 
MD & 740-860 & 2.1 & 4.3 & 2.4 & 4.1 & 4.4 & 2.1 & 2.3 & 3.1 & 2.7 & 1.0 & 1.2 & 0.4 \\
MD & 740-860 & 0.1 & 0.2 & 2.0 & 0.1 & 0.3 & 0.4 & 0.2 & 1.1 & 0.9 & 0.2 & 0.0 &  \\
\noalign{\smallskip} 
MDB & 740-860 & 3.3 & 3.9 & 3.7 & 3.4 & 3.8 & 2.9 & 2.6 & 3.4 & 3.4 & 1.5 & 1.7 & 0.5 \\
MDB & 740-860 & 0.1 & 0.1 & 0.1 & 0.1 & 0.1 & 0.1 & 0.1 & 0.5 & 0.5 & 0.1 & 0.1 &  \\
\noalign{\smallskip} 
\end{tabular}
\end{flushleft}
\end{table*}

\section{Quantifying the peanut shape }
\label{sec:peanut}

\begin{figure*}
\includegraphics{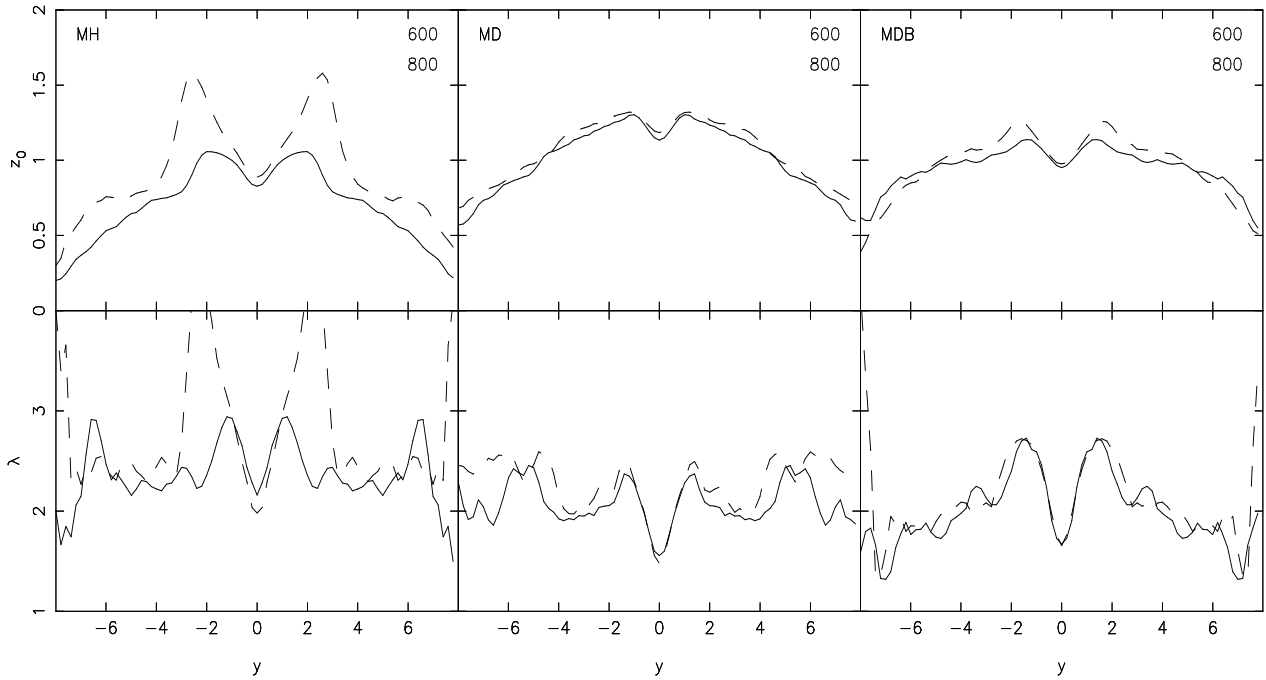}
\vspace{9.3cm}
\caption{
Parameters $z_0$ (upper panels) and $\lambda$ (lower panels) of the
generalised gaussian fitting best cuts parallel to the $z$ axis. The
left panels correspond to model MH, the middle one to model MD and the
right one to model MDB. The name of the simulation is
given in the upper left corner and the times in the upper right corner
of the upper panels. The solid line corresponds to time 600 and the
dashed one to time 800.
}
\label{fig:z0l}
\end{figure*}

From the third row of panels of Figures~\ref{fig:basic600} and
\ref{fig:basic800} we can see that simulation MDB and the earlier time
of simulation MH show a clear peanut shape, simulation MD is boxy,
while in the later times simulation MH has clear X signature. In this
section we will 
make this comparison more quantitative. We explore two 
different ways of quantifying the peanut shape.

The first relies on cuts parallel to the major axis when the galaxy 
is observed side-on, as introduced in section~\ref{sec:edgeon}.
From such cuts the strength of the peanut can be parametrised by

\begin{equation}
SP_1 = \Sigma_{max} (y, z_{ref}) / \Sigma (0, z_{ref}),
\end{equation}

\noindent
where $y$ the distance from the center along the cut, $z_{ref}$ is the $z$
value at which this cut is made, $\Sigma (y, z_{ref})$ is the projected
density along the cut, $\Sigma_{max}$ is the maximum value of the density
along the cut and $\Sigma (0, z_{ref})$ is the value at the center of the cut. 
The value of 
$SP_1$ is heavily dependent on the value of $z_{ref}$. For too low a 
value of $z_{ref}$ the peanut strength is not well revealed, while for too
high a value $\Sigma (0, z_{ref})$ = 0, and therefore the above definition 
is not applicable. We have 
chosen $z_{ref}$ = 1, which we believe is a reasonable compromise.
Such cuts can also tell us the radial size of the peanut, defined as the
value of $y$ for which $\Sigma (y, z_{ref})$ is maximum. This  
also depends on $z_{ref}$. It increases strongly with $z$ for
small values of $z$ and less so for larger $z$ values. 
We choose as our first measure of the
peanut length $L_{P1}$, the value of $y$ for which $\Sigma (y, 1)$ is maximum.

Applying this definition to model MH we get $SP_1$~=~5.5 for time 600 
and $SP_1$~=~12 for time 800. The corresponding lengths of the peanut 
are $LP_1$~=~1.7 and $LP_1$~=~2.2 correspondingly. For model MDB we
get for  $SP_1$ and time 600 (800) the value of 2.2 (3.0), and for
model MD the value 1.5 (1.2). For the latter model the minima and
maxima, and therefore the values of the peanut length, are 
poorly defined, showing that the form is more a box than a peanut.

For the second method we ``observe'' our
models side-on and make cuts parallel to the 
$z$ axis at different values of $y$. For each cut we make a profile 
of the projected surface density as a function of $z$. We symmetrise
the profiles with respect to $z$~=~0, fit
to them generalised gaussians of the form 
$ exp (- (z/z_0) ^ {\lambda}) $ and thus 
determine the values of $z_0$ and $\lambda$ for which the generalised
gaussian fits best the profile. Larger
values of $z_0$ correspond to broader gaussians, while the parameter
$\lambda$ defines the shape of the generalised gaussian. For 
$\lambda$ = 1 the generalised ellipse becomes an exponential and
for $\lambda$ = 2 a standard gaussian. For small values of $\lambda$ 
the generalised gaussian is very peaked at the center and for large
ones it has a relatively flat top.

Figure~\ref{fig:z0l} shows the values of $z_0$ and $\lambda$ as a
function of the $y$ value at which the cut was made. Since the results
are relatively noisy we apply a sliding means. For model MH $z_0$ 
shows a clear minimum at 
$y$ = 0 followed by a clear maximum. This is a  
signature of a peanut. Thus our second way of quantifying the peanut 
is from the ratio of the
maximum of $z_0$ to its value at the center, namely 

\begin{equation}
SP_2 = z_{0,max}/z_0(0).
\end{equation}

\noindent
This is 1.3 (1.8) for times 600 (800) and model MH. 
For model MD, the value of $SP_2$ at time 600 (800) is
1.1 (1.1), and for model MDB 1.2 (1.3). From these and a number of
other cases we can see that $SP_2$ can distinguish well between shapes
which are boxy, where it gives values hardly above 1, peanuts, where
it gives larger values, and X shapes, where it gives considerably larger
values. It can thus be used as a measure of the box/peanut/X strength.

The location of the maximum of $z_0$ can be used as a measure of the
radial extent  
of the peanut ($LP_2$). We find $LP_2$~=~2.0
(2.6) for time 600 (800). 
The results for the peanut length obtained with the two above definitions
averaged in the [540, 660] and [740, 860] time ranges are listed in
columns 13 and 14 of Table~\ref{tab:barlength}. The agreement between the
two results is of the order of 10 per cent,
except for the later time of model MD, where the difference is 20 per
cent. We can thus consider either of them, or their average, as a reliable
estimate of the peanut length. They show clearly that the radial
extent of the peanut increases considerably with time. We also
calculated the ratio of the peanut length (as the average of the two
estimates) to the bar length ($L_2$) and give it in the last column of
Table 1. 

The parameter $\lambda$ also shows a minimum in the center surrounded
by two maxima, one on either side. This does not necessarily imply
that it can be used as a measure of the peanut strength or
length. Indeed the shape of the gaussians need not correlate with
their width.

A different parametrisation, based on the $a_4$ parameter (Bender \tal
1989)
often used for elliptical galaxies, will be discussed in a future
paper in collaboration with M. Bureau.

\section{Rotation }
\label{sec:rot}

\begin{figure*}
\includegraphics{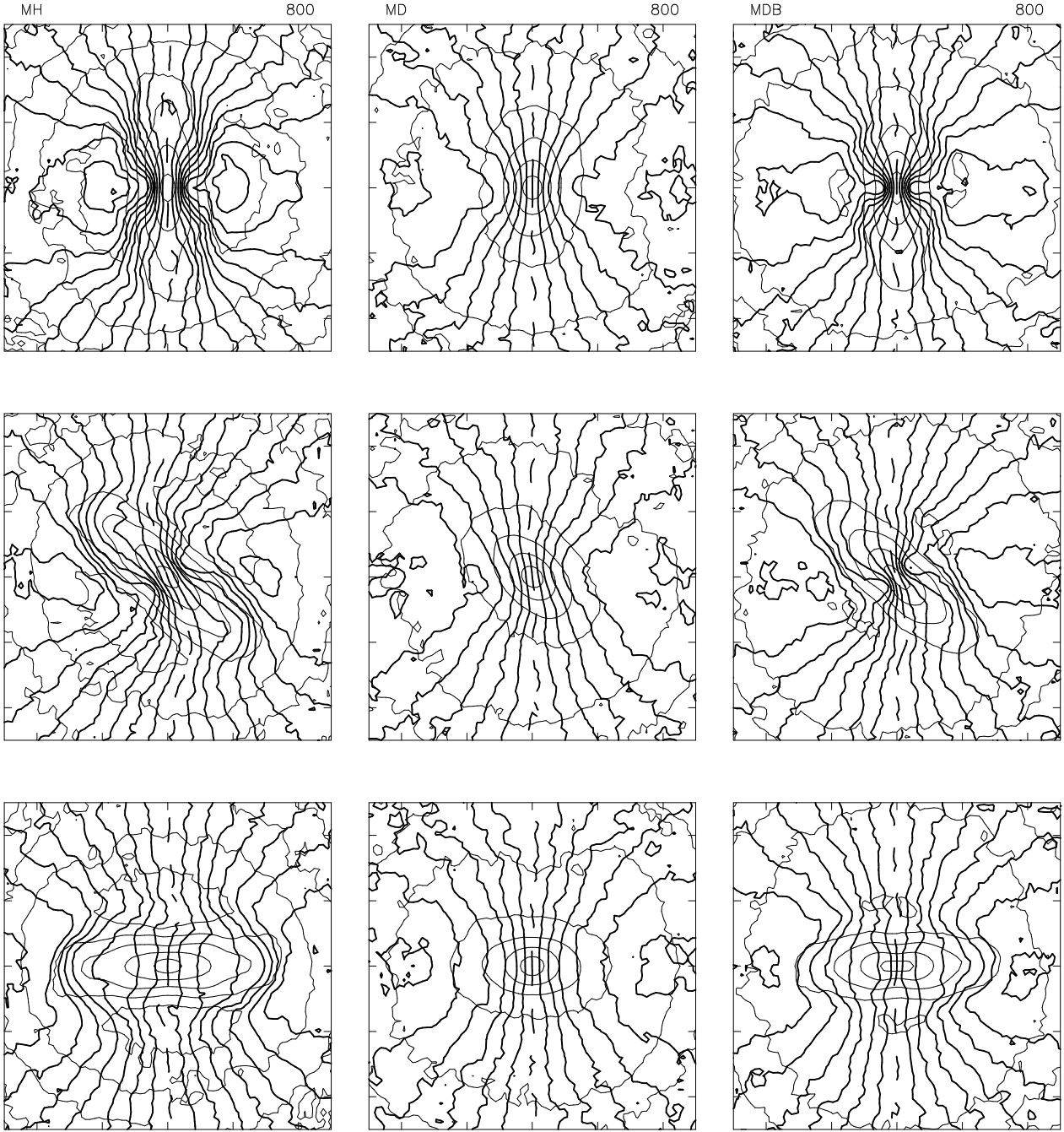}
\vspace{20.0cm}
\caption{
Velocity field of the disc, for three different
orientations of the bar: along the $y$ axis (upper panels), at 45 degrees to
it (middle panels), and along the $x$ axis (lower panels). The
isovelocities are given by thick lines and the kinematic major axis, 
i.e. the isovelocity corresponding to the
systemic velocity (in our case zero), with a dashed line. The 
$\Delta v$ between two consecutive
isovelocities is 0.1. We also overlay the isodensities with thin
line contours. The line of
sight is along the $y$ axis. The left panels correspond to model MH,
the middle ones to model MD and the right ones to model MDB, all
at time 800.}
\label{fig:faceonv}
\end{figure*}

The orbits of particles in a barred galaxy are far from circular, and
this of course reflects itself on the galaxy velocity
field. 
Figure~\ref{fig:faceonv}  shows the velocity fields of our three fiducial 
models. It is obtained in a way that varies somewhat from that of
observations, yet it is the most convenient for comparing with
observations of galaxies at intermediate inclinations, for which the
contribution of the $z$ component of the velocity is relatively small.
To obtain it we project all particles on the equatorial plane, 
observe their $v_y$ velocity component and plot the corresponding
isovelocities. For the three
views shown in the figure this is equivalent to observing along the bar major
axis, at 45 degrees to it and along the bar minor axis, respectively.

Model MD gives velocity fields analogous to those given by 
previously published models. When we observe along the bar major axis,
the isovelocities show a
characteristic concentration towards the central region, 
due to the fact that particle orbits are elongated
along the bar and the velocity along an orbit
is larger at pericenter than at apocenter. The intermediate angle velocity
field shows the $Z$ structure characteristic of barred galaxy velocity
fields (see e.g. Peterson \tal 1978, for NGC 5383), and finally the velocity
field obtained when we view along the bar minor axis shows a sizeable
area of solid body rotation in the inner parts.

Several of these
features are seen also in the case of model MH, but with some notable
differences. Thus when we observe along the bar major axis we note a 
strong pinching of the isovelocities in the innermost region, on or 
near the bar minor axis. For the 45 degrees
orientation the $Z$ shape is much more pronounced than in the MD case,
which could be expected since the bar in model MH is so much
stronger. The greatest surprise, however, comes from the last
orientation, where we view the disc along the bar minor axis. The innermost
solid body rotation part is there, as for model MD. But as we move
away from the kinematical minor axis the isovelocities show a clear
wavy pattern, indicating that the mean velocities is lower at the 
ends of the bar than right above or right below that. This is due to the
fact that near the ends of the bar the particles are at their apocenters.
The mean velocities in those regions can be further lowered
if the corresponding periodic orbits have loops at their apocenters. 
Athanassoula (1992) discusses such loops, and the regions
where they appear in her models correspond roughly to the low velocity
regions discussed here. She also shows that such loops occur mainly in 
periodic orbits in strong bars, i.e. that such loops are more liable 
to be found in model MH than in model MD. A more quantitative
comparison will have to wait for a complete study of the orbital
structure in these models, which should furthermore elucidate the
formation and properties of the ansae. Let us also note here that
the velocity field is that of a stellar component and should not be
compared to those obtained by observing the gas or from
hydrodynamical simulations.

The velocity field of model MDB is
intermediate of those of models MH and MD. This holds both for
the $Z$ pattern seen in the 45 degrees orientation and the
wavy pattern at the ends of the bar when we view along the
bar minor axis.  
The central crowding, however, is more important than in models MD 
and MH. This is due to the extra central concentration of the bulge component.
 
Let us note in passing that no velocity gradients are seen along the 
line of sight when that coincides with the major or the minor axis 
of the bar. Furthermore, hardly any velocity gradient can be seen
in the MD and MDB cases, even when the line of sight is at 45 
degrees to these axes. Thus the existence of a velocity gradient along
the minor axis is not a very good criterion
to picking out bars and oval distortions and should be left for 
cases where only limited information along slits is available.
If a 2D velocity field is available one should rather look at 
whether the kinematical major and minor axes are orthogonal to
each other (cf. e.g. Bosma 1981 for a discussion) and at the
twists of the isovelocity contours.

\begin{figure*}
\includegraphics{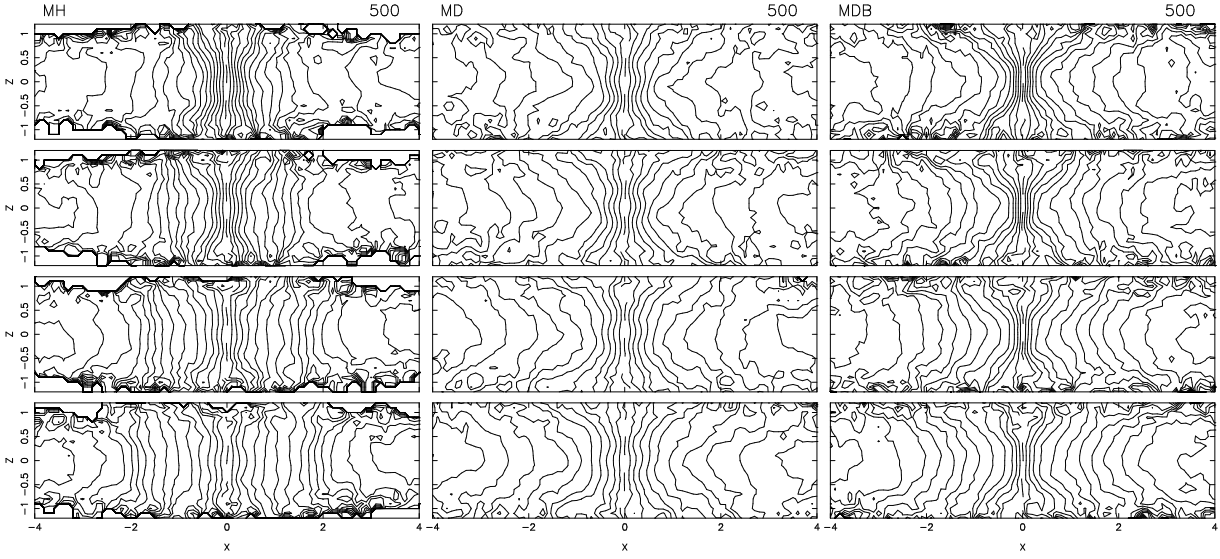}
\vspace{9.5cm}
\caption{
Velocity field of the three fiducial discs seen edge-on, for four different
orientations of the bar: end-on (upper row), at 30 degrees to
the line of sight (second row), at 60 degrees to
the line of sight (third row) and side-on (fourth row).
The kinematic major axis is given by a dashed
line. The left panels correspond to model MH,
the middle ones to model MD and the right ones to model MDB, all
at time 500.} 
\label{fig:edgeonv}
\end{figure*}

Figure~\ref{fig:edgeonv} shows the velocity
field that is obtained when the disc is seen edge-on, for four
different angles of the bar with respect to the line of sight, for
models MH, MD and MDB respectively. Again there are important 
differences between models MH and MD. For
model MH the mean rotation does not depend on the the distance $z$ from
the equatorial plane in a central region, whose
size depends on the orientation of the bar. Perpendicularly to the
equatorial plane it extends roughly as far as the bar material
extends. The distance along the equatorial plane is smallest when
the bar is seen end-on and biggest when the bar is seen side-on. In
the latter case it extends over most of the area covered by
the peanut. On the other hand in the MD model the mean rotational 
velocity drops considerably with increasing distance from the equatorial 
plane, thus again providing a clear dichotomy between the
two fiducial cases. Model MDB has a velocity field similar to that of model MD.

As already discussed, the form of the MH bar viewed side-on evolves from 
box-like to peanut and then to 
X shaped. When it reaches this last stage the area within which the
rotation does not depend on $z$
is somewhat less extended, particularly in the low density
areas on and around the $z$-axis, on either side of the center. 

\section{Velocity dispersions of a thin strip of particles taken along
the bar major axis }
\label{sec:disp}

\begin{figure*}
\includegraphics{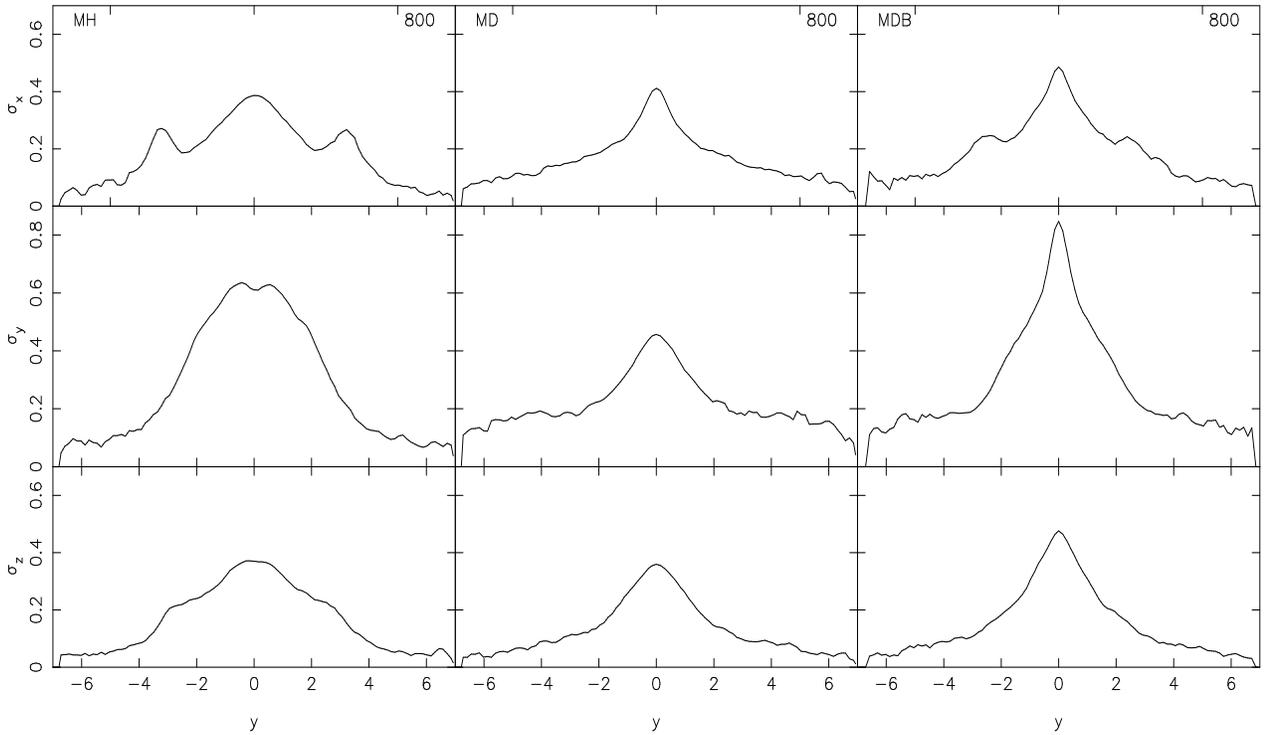}
\vspace{10.cm}
\caption{
Velocity dispersion as a function of distance from the center for our
three fiducial models at time 800, as discussed in
section~\ref{sec:disp}. The components $\sigma_x$, $\sigma_y$ and 
$\sigma_z$ are shown in the upper, middle and lower panels
respectively. The
left panels corresponds to model MH, the middle ones to model MD and the
right ones to model MDB. The simulation name is given in the upper
left corner and the time in the upper right corner of the upper panels.
}
\label{fig:vdisp}
\end{figure*}

In order to get more information on the motions in our fiducial models
we have isolated a strip of particles centered on the 
bar major axis and having a width of 0.07. 
Figure~\ref{fig:vdisp} shows the three components of the velocity
dispersion, $\sigma_x$, $\sigma_y$ and  $\sigma_z$, as a function of
the distance from the center of the 
galaxy measured along the strip, at time 
800. The $\sigma_x$ for model MH shows two
sizeable maxima, one at either side of the center. They should be
due to the form of the $x_1$ orbits in that region, either to the
loops that these orbits can have on the major axis, or to their more
rectangular like shape, and will be discussed in those terms in
a future paper, where the orbital structure of these models will be
presented. Here we will note that the shape of the isophotes in
that region is very rectangular-like for model MH.  
The $\sigma_x$ for model MDB also shows two similar maxima, of
relatively lower amplitude. This model also has bar isophotes with
a rectangular-like shape, but the maximum rectangularity occurs at
larger distances from the center than the secondary maxima of the
$\sigma_x$ profile. No such maxima can be seen for model MD. 

The central value of all three components for model MDB are much
larger than those of the other two models, presumably due to the 
presence of the bulge. This is particularly clear for the $\sigma_y$
component. 
Also the velocity dispersions are large over a more extended
region in model MH than in model MD. This can be easily
understood by the fact that the size of the bar is larger, and that
the velocity dispersions are larger within the bar region.

It is also worth noting that the $\sigma_y$ component for model MH has
a minimum at the center, surrounded by two low maxima close to it and
on either side of the bar, 
and followed by a quasi-linear drop. This is not the case for models MD and
MDB, where the maximum velocity dispersion is reached at the
center. Again the explanation of this in terms of orbits will be
discussed after we present the orbital structure in each model.

\section{The bulge }
\label{sec:bulge}

We used the inertia tensor to obtain information on the shape of the
bulge. For this we first calculated the local density at the position
of each bulge particle, using the distance to its six nearest
neighbours (Casertano \& Hut 1985) and then sorted the bulge particles
in order of increasing local density. 
We believe that the density is more appropriate than the radius --
which introduces a 
circular bias -- or than the binding energy -- which also introduces
such a bias, albeit much less than the radius. We then
discarded the 10 per cent in the most dense environment, part of which could
be influenced by the softening, and the 10 per cent in the least dense
environment, which contains particles at very
large distances from the center, divided the
remaining particles into five groups of equal mass and calculated the
eigenvalues and eigenvectors of the inertia tensor separately for each
group and then for the five groups together. The axial ratios can then be
obtained (e.g. Barnes 1992) as those of the homogeneous ellipsoid that
has the same moment of inertia. We thus obtain 

{\smallskip}
\noindent
$b_b/a_b = \sqrt{q_2/q_1}$

{\smallskip}
\noindent
and 

{\smallskip}
\noindent
$c_b/a_b = \sqrt{q_3/q_1}$

{\smallskip}
\noindent
where $a_b$, $b_b$ and $c_b$ are the lengths of the three principal
semi-axes of the bulge, and $q_1$, $q_2$ and $q_3$ are the eigenvalues 
of the inertia tensor. 

When considered as a whole, the bulge of the fiducial MDB model is an
oblate object, with the 
shortest dimension along the $z$ axis, i.e. perpendicular to the disc
plane. The value of $c_b/a_b$ is around 0.9. However, when we consider
each group of particles separately, the departure from sphericity can
be much more 
important. Thus the group with the highest densities, which has a mean
distance from the center of 0.3, is somewhat triaxial with axial
ratios roughly 0.75 and 0.7 respectively. The median group, with a
mean distance 
from the center of 1.2, has axial ratios of roughly 0.9 and 0.8
respectively. In 
general, as we go from highest to lowest densities, the shape becomes
gradually more spherical.

\section{Summary and discussion}
\label{sec:summary}

In this paper we present three simulations of bar unstable disc galaxies.
Initially in the first model, MH, the halo mass is concentrated in the
inner parts, while
in the second one, MD, it is the disc mass that dominates in the 
central parts. The third model, MDB, is similar to model MD, but has 
also a bulge of mass 60 per cent of the disc mass. In all three cases the 
halo mass is about five times the mass of the disc. All three models 
evolve and form a bar in their central parts. The properties of the 
bar, however, are very different in the three cases.

Model MH forms a very strong bar, which is long and thin when viewed 
face-on. It is surrounded by a massive inner ring, slightly elongated 
along the bar. Seen edge-on the galaxy has a peanut or X shape. This
model has initially no bulge, so the particles forming the peanut 
are disc particles, or, more precisely, bar particles. It is thus
not appropriate to call this component a bulge, as is often done. This is not,
however, the only way that we could have attributed material to a 
bulge in this bulge-less galaxy. When seen end-on the disc material
has a definite central spheroidal concentration, extending well out
of the plane, which could easily be mistaken as a bulge sticking out 
of an edge-on disc galaxy. Surface photometry will only enhance 
this impression. Projected density profiles, obtained both from 
face-on and edge-on views, show clear central concentrations in the
inner part, which would again easily be attributed to a bulge.

The bar of model MH is not only strong and long, but it is also thin 
and rectangular-like, as we could show by fitting generalised ellipses 
to the isophotes. The rectangularity is particularly strong in the 
outer parts of the bar. Within the main bar region the ellipticity of the bar
isophotes does not change much with distance from the center,
and then it drops abruptly to near-circular in the ring
region. Outside 
the ring the isophotes become somewhat more elliptical shaped and 
are oriented perpendicular to the bar. At sufficiently large distances
from the center they are again near-circular. Thus even in this very
strongly barred case 
the ellipticity of the outer disc isophotes is sufficiently low 
for observers to be able to consider them circular and use them for
deprojecting.

For model MH the projected surface density profiles, obtained from cuts
along the bar major axis of the galaxy viewed face-on, show a flat
region on either side of the  
central concentration, the end of which can be used as a measure of
the bar length. Similar profiles, now from the galaxy seen edge-on,
also show this characteristic ledge. Cuts which are offset from the
equatorial plane and parallel to the bar major axis
reveal a two-horned shape, characteristic of a peanut. They
can be used to parametrise the length and the strength of the peanut.

We also Fourier analysed the face-on density distribution of the disc
particles. We find that the $m$~=~2 component is quite large, between
0.6 and 0.8 for the times we discuss here. The position 
of the maximum is well within the bar; half or a quarter of the way
to the center. Nevertheless the $m$~=~2 component is important 
all through the bar region. The relative amplitudes of the higher
$m$ components are smaller than that of the $m$~=~2, but are 
still very big. Even  $m$~=~8 is of the order of 
one third of the $m$~=~2. The distance of the location of the 
maximum of the relative amplitude from the center increases 
with $m$.

The velocity field also shows very strong deviations from circular
motion, due to the presence of the bar. When the line of sight 
is along the bar major axis, or at 45 degrees to it we observe
crowding of the isovelocities in the center-most areas.
The 45 degrees orientation shows very clearly the strong 
$Z$-type isovelocities, classical of barred galaxies. When the 
line of sight is along the bar minor axis the isovelocities 
passing near the ends of the bar show a strong wavy form. 
This is due to the low mean velocities in the end of the bar region.

When viewed edge-on the model exhibits strong cylindrical rotation
over a large area. The bar signature is also clear in the 
velocity dispersions, which are high in all the bar region. 
The component perpendicular to
the bar major axis shows clear local maxima around the ends
of the bar. The component parallel to the bar major axis shows a
shallow minimum or a plateau at the center.

The bar in the MD model is quite different. It is considerably
shorter, thicker and less rectangular-like than the MH bar. 
Viewed edge-on it has a form which is better described as 
boxy. Its projected density profiles decrease steeply with radius in the
bar region and that both in the face-on and edge-on view,
as opposed to the flatter profiles of the MH bar. The relative 
amplitude of the $m$~=~2 component is less than 0.4, and 
the $m$~=~6 and 8 components are negligible. The relative $m$~=~4
component is considerably smaller than the relative 
$m$~=~8 component of the MH model. Viewed edge-on model MD
does not display cylindrical rotation.

Finally, model MDB has a bar which is intermediate in length
and shape to that of the previously discussed two models.
Viewed edge-on it has a peanut shape, which never evolves to an X shape,
at least within the times considered here. 
Its projected density profiles, both for the face-on and
the edge-on viewing, as well as its velocity field are intermediate of
those of models MH and MD. In the face-on profiles there is more
difference between the major and the minor axis profiles than in the
MD case. On the other
hand there is no flat part, except perhaps at the very late times of
the evolution, where such a structure starts to form. 

In all three cases the formation and evolution of the bar is followed 
by a substantial inflow of disc material towards the central parts.
This implies qualitative changes for model MH. Indeed at the 
initial times the halo contribution is somewhat larger than the disc
one in the 
inner parts. The halo density distribution does not change 
much with time, while the disc becomes considerably more centrally 
concentrated. Thus at latter times the disc dominates within the 
central region, to a distance larger than an initial disc scale length. 
For the MD and MDB spheroids also the density distribution does not 
change much with time, and again the disc component becomes more 
centrally concentrated. Thus the central areas become even more 
disc dominated than they were at the beginning of the simulation.
The difference, however, is quantitative, rather than qualitative, 
as it was for the MH model.

We used a number of different ways of measuring the bar length. 
Some of them are more suited for MH type models, others for MD types,
while others can be used for both. Unfortunately we could not find any
criterion which could do well for all simulations and all
times. Average values give satisfactory estimates within their error
bars, but their error bars are rather large. This will prove to be a
major problem when we will want to compare the length of the bar to
the corotation radius, to see how evolution affects this ratio and
whether it stays compatible with observational limits (cf. Debattista
\& Sellwood 2000).

We also introduced two ways of measuring the strength and
length of the peanut. They allow to distinguish between boxy,
peanut-shaped and X-shaped outlines. They also show that the length of
the peanut is considerably shorter than the length of the bar.

\parindent=0pt
\def\rr{\par\noindent\parshape=2 0cm 8cm 1cm 7cm}
\vskip 0.7cm plus .5cm minus .5cm

{\Large \bf Acknowledgments.} We would like to thank A. Bosma for useful
discussions and J.~C. Lambert for his help with the GRAPE
software and the administration of the simulations. We
would also like to thank the IGRAP, the Region PACA, the
INSU/CNRS and the University of Aix-Marseille I for funds to develop
the GRAPE computing facilities used for the simulations discussed in
this paper. Part of this paper was written while one of us (E.A.) 
was visiting INAOE. She would like to thank the ECOS-Nord and the ANUIES for
financing this trip and INAOE for their kind hospitality.
\vskip 0.5cm

{\Large \bf References.}
\vskip 0.5cm
\parskip=0pt
\rr{Aguerri J.A.L., Mu\~noz-Tu\~n\'on C., Varela A.M., Prieto
M., 2000, \AAA, 361, 841}
\rr{Athanassoula E., 1992, \MN, 259, 328}
\rr{Athanassoula E., Bosma A., Lambert, J.C. \& Makino J., 1998, \MN,
293, 369}
\rr{Athanassoula E., Morin, S., Wozniak, H., Puy D., Pierce M.,
Lombard J. \& Bosma A., 1990, \MN, 245, 130}
\rr{Bosma A., 1981 \AJ 86 1825}
\rr{Barnes J. E., 1992, \ApJ, 393, 484}
\rr{Bender R., Surma P., D\"obereiner S., M\"ollenhoff C., Madejsky R., 1989,
\AAA, 217, 35}
\rr{Casertano, S., Hut, P., 1985 ,\ApJ, 298, 80}
\rr{Debattista V.P., Sellwood J.A., 2000, \ApJ, 543, 704}
\rr{Hernquist L., 1993, \ApJS, 86, 389} 
\rr{Kawai, A., Fukushige, T., Makino, J., Taiji, M., 2000, \PASJ, 52, 659}
\rr{Kawai, A., Fukushige, T., Taiji, M., Makino, J., Sugimoto, D., 1997,
\PASJ, 49, 607}
\rr{Kent, S., 1986, \AJ, 91, 1301}
\rr{Kormendy J., 1983, \ApJ, 275, 529}
\rr{L\"utticke, R., Dettmar, R.-J., Pohlen, M. 2000, \AAA, 362, 435}
\rr{Ohta K., Hamabe M., Wakamatsu K., 1990, \ApJ, 357, 71}
\rr{Peletier R.F., Balcells M., 1996, \AJ, 11, 2238} 
\rr{Peterson C.J., Rubin, V.C., Ford, W.K., Thonnard, 1978, \ApJ, 219, 31}
\rr{Toomre A., 1964, \ApJ, 139, 1217}

\appendix

\section[]{Initial conditions}
\label{sec:initdetail}
In order to generate the initial conditions of our simulations we
widely followed
the method described by Hernquist (1993), with the following small
differences: 
\begin {enumerate}
\item
For the radial velocity dispersion Hernquist (1993) adopts
$\sigma_R^2(R)=C exp(-R/h)$, where the constant $C$ is 
normalized so that the Toomre Q parameter (Toomre 1964) has 
a prescribed value at a
given radius. One technical problem with this choice is that the central parts
may be very hot, making the epicyclic approximation, on which the
calculation of the asymmetric drift and of the azimuthal velocity
dispersion is based, totally inadequate. A second problem is that, for
certain choices of $C$ and of the reference radius, the disc may turn out to
be locally unstable at certain radii. As an alternative we have
adopted a $Q$ that is constant all through the disc. This also has
some technical problems, but we found them to be less severe than
those of Hernquist's choice. Namely in cases with strong bulges one
has to take care that the forces are properly calculated in the
central parts of the disc before obtaining the epicyclic frequency
(see also point iii below). Even so, in difficult cases, it is possible that
the streaming velocity becomes larger than the circular velocity. In
such cases we artificially lower the value of the streaming velocity
to the value of the circular velocity.
\item
In Hernquist's method, the velocities of the halo particles are drawn
from a gaussian whose second moment is the velocity dispersion of the
halo distribution at the radius under consideration. While a gaussian is the most
natural choice, it has the inconvenience of extending to
infinity. Thus particles are often drawn with velocities larger than
the escape velocity. To avoid this we have used another function,
namely~:

\[ F(v,r)=\left\{ \begin{array}{ll}
C_1\left(1-\frac{v_i^2}{C_2^2}\right) & \mbox{$v_i \leq C_2$} \\
0 & \mbox{$v_i > C_2$} \\
\end{array}
\right. \]

\noindent
where $v_i$ is the $x$, $y$, or $z$ component of the velocity and the
constants $C_1$ and $C_2$ are calculated so that the zeroth and second
moments are the same as those of the gaussian. 
This function limits the values of the velocity components to the
interval $(-C_2,C_2)$. 
While it is still possible for some particles to be drawn with
velocities larger than the escape
velocity, their number is significantly smaller
than in the case of the gaussian.
\item
In order to calculate the epicyclic frequency we need first to 
calculate the force on the disc particles, and for that we used 
direct summation,
including the softening. This allows a more accurate determination of
the forces,
particularly in the central parts of models with bulges. In fact
Hernquist (1992) does not give sufficient information on how he
calculates the force, so we can not be sure that this is indeed a
difference between his method and ours.

\end{enumerate}

\section[]{Measuring the mass, radius and axial ratio of the ring}
\label{sec:techniques}
We developed two different, independent, methods of fitting a ring 
and obtaining 
its basic parameters. In the first we fit a single function $I(R, \theta)$
simultaneously to most of the face of the galaxy. We first exclude the 
regions with $R < 1$, as they do not contain information relative to the 
ring, and the regions where $cos (\Delta \phi) < 0.5$, where $\Delta \phi$ 
the azimuthal angle measured from the major axis of the bar, since there the 
ring does not detach itself sufficiently well from the inner part of the 
profile to allow an accurate fit. To the remaining part we fit a function

{\medskip}
\noindent
$
I(R,\phi)=C_0(\phi)~R^{-C_1(\phi)} + \\
~~~~~~~~~~~~~~C_2(\phi)~exp(-(R-C_3(\phi))^2/C_4(\phi)^2)
$
,

{\medskip}
\noindent
where $C_i(\phi) = P_{2i} + P_{2i+1} cos(2\phi)$, $i$ = 0,...4.
This is a 10 parameter fit and we call this method and the
corresponding fit ``global''.  

We also used a different method, 
inspired from photometric work on barred galaxies. 
We first obtained a face-on projected density of the model and then
made 100 radial cuts 
with an angle of 3.6 degrees between two consecutive cuts. For each cut
we make a two component fit of the radial density profile using the
functional form 

{\smallskip}
\noindent
$
I(R,\phi)= D_0~R^{-D_1}~+~D_2~exp [-((R-D_3)/D_4)^2]
\label{eq:ring1}
$,

{\smallskip}
\noindent
$D_0, 
D_1, D_2, D_3$ and $D_4$ are constants and are determined independently for
each radial cut, i.e. for each value of the azimuthal angle. 
When doing these fits we do not take into account the regions with
$R <$ 1, since these are too near the center to contain any useful
information about the ring. Since this method deals with each radial 
cut independent of all the others we call it ``local''.

We thus have for each radial cut a set of $D_i$ values, $i$~=~0,..,4. 
We now discard
those cuts - and the corresponding values of $D_i$ - which are at
angles less than 30 degrees from the bar major axis, as well as all
other cuts where the ring contribution does not detach itself from the
background. 
To the remaining values of ($D_i, \phi$) we then fit the simple forms
\begin{equation}
D_i (\phi) = p_i + q_i cos (2 \phi),~~~~~~i=0,..,4
\label{eq:ring2}
\end{equation}
\noindent
where $p_i$ and $q_i$ are constants. 

In the above two methods we have parametrised all the density 
except for the ring by a
simple power law. It could be argued that it would have been more
realistic had we used an exponential disc plus a functional form to
describe the bar. This, however, would have implied a very large
number of free parameters 
and made the problem badly determined. We find that in most cases a
gaussian profile gives a very good fit to the 
ring component, while the power law profile, is adequate for the
remaining part. In this way we limit the number of free parameters.
In a large number of cases we checked by eye that the results are
satisfactory. As a further measure of the quality of the  
fits we could compare how well the estimates given independently 
by the two methods agree. In general we find that the radius of the
ring is very well determined. This is less so for its width and mass,
since they can be considerably influenced by the wings of the gaussian. This is
not too serious for narrow rings, but much more so for wide ones.

\label{lastpage}

\end{document}